\documentclass[aip,reprint,floatfix,onecolumn]{revtex4-1}
\usepackage{amsfonts,amssymb,amsmath,array}
\usepackage{bm}
\usepackage{cancel}
\usepackage[final]{graphicx}
\usepackage{epstopdf}
\usepackage[usenames]{color}
\usepackage{xcolor}
\usepackage{soul}
\usepackage{color}
\definecolor{ForestGreen}{RGB}{34,139,34}   % new changes (November 2026)
\DeclareMathOperator{\Tr}{Tr}

\newif\ifnotes
\notestrue     % change to \notestrue to show the notes

\begin{document}

\title{Magnetic tuning of odd elastic instabilities in chiral active crystals}
\author{U. Marini Bettolo Marconi}
\affiliation{School of Sciences and Technology, University of Camerino, Via Madonna delle Carceri, Camerino, Italy}
\author{L. Caprini}
\affiliation{Sapienza University of Rome, P.le A. Moro 2, Rome, Italy}

\newcommand{\beq}{\begin{equation}}
\newcommand{\eeq}{\end{equation}}
\newcommand{\bea}{\begin{eqnarray}}
\newcommand{\eea}{\end{eqnarray}}
\newcommand{\br}{{\bf r}}
\newcommand{\rbu}{{\bf u}}
\newcommand{\bu}{\hat{\bf u}}
\newcommand{\bv}{{\bf v}}
\newcommand{\bR}{{\bf R}}
\newcommand{\bRz}{{\bf R}^0}
\newcommand{\bk}{{ \bf k}}
\newcommand{\bx}{{ \bf x}}
\newcommand{\vv}{{\bf v}}
\newcommand{\vvt}{\tilde{\bf v}}
\newcommand{\bn}{{\bf n}}
\newcommand{\mb}{{\bf m}}
\newcommand{\bq}{{\bf q}}
\newcommand{\bK}{{\bf K}}
\newcommand{\rb}{{\bar r}}
\newcommand{\rr}{{\bf r}}
\newcommand{\eb}{{\bf e}}
\newcommand{\tbfF}{\tilde{\bf F}}
\newcommand{\bfnabla}{\boldsymbol{\nabla}}
\newcommand{\kk}{\boldsymbol{\kappa}}
\newcommand{\greeketabold}{\boldsymbol{\eta}}
\newcommand{\xxi}{\boldsymbol{\xi}}
\newcommand{\cchi}{\boldsymbol{\chi}}
\newcommand{\bomega}{\boldsymbol{\Omega}}
\newcommand{\smallomega}{\boldsymbol{\omega}}
\date{\today}

\begin{abstract}
We investigate the dynamical properties of an odd active crystal whose elementary units follow an inertial dynamics and experience a Lorentz force induced by an external magnetic field. The continuous injection of energy associated with active chirality gives rise to odd elasticity, manifested through transverse forces between the crystal constituents.
We show that the interplay between the magnetic field and these transverse forces produces a distinctive long-wavelength instability at vanishing wavevector, which is absent in both non-magnetic elastic crystals and passive crystals subjected to a magnetic field. Moreover, the magnetic field provides a means of tuning the collective oscillations of the crystal, enabling the emergence of exceptional points and tuning the dynamical instability induced by odd elasticity at finite wavevectors.
Our analytical results yield explicit predictions for the displacement spectrum, static structure factor, and intermediate scattering function. These findings suggest that magnetic fields could provide a powerful tool for controlling collective phenomena in high-density systems of granular or colloidal spinners.

\end{abstract}

\maketitle

\section{Introduction}

Chiral active matter \cite{lowen2016chirality, liebchen2022chiral, caprini2026modeling} encompasses a broad class of non-equilibrium systems in which the continuous consumption of energy drives rotational motion \cite{van2008dynamics,caprini2019active,hargus2021odd} at the single-particle level -- as for chiral colloids that spin because of a non-rotationally symmetric body \cite{ebbens2010self, kummel2013circular, narinder2018memory} or an external magnetic field~\cite{yan2014rotating,massana2021arrested} -- or, upon coarse-graining, effective non-conservative interactions between particles \cite{caprini2026modeling}. Chirality can then produce transverse forces that are not collinear with the line connecting interacting particles~\cite{berry2012classical,caprini2025bubble,guo2025diffusion}, arising through hydrodynamic interactions between rotating colloids~\cite{shen2023collective} or through rotational friction in collisions between granular spinners~\cite{caprini2026modeling,digregorio2026phase}.

At the collective level, chirality is known to suppress clustering \cite{ma2022dynamical,bickmann2022analytical,lei2023collective,langford2025phase} and promote micro-phase separation~\cite{semwal2024macro}, to generate vortex-like flow patterns \cite{shee2024emergent} together with a spontaneous emergence of angular momentum in dense systems~\cite{marconi2025spontaneous}. When chirality specifically gives rise to transverse (odd) interactions, it produces edge currents in a cluster phase \cite{caporusso2024phase} and a qualitatively distinct phase transition from a homogeneous to an inhomogeneous, bubble-patterned phase~\cite{caprini2025bubble,guo2026tuning}, termed BIO~\cite{caprini2025bubble}, observed both in rotors immersed in a fluid and subject to hydrodynamic interactions~\cite{shen2023collective} and in granular-spinner models~\cite{digregorio2026phase,carenza2026phase}. This phase transition has been recently rationalized through a hydrodynamic approach based on odd viscosity and an additional torque density~\cite{marini2026emergent,maire2026kinetic}.

When such interactions act in a crystalline environment, they naturally lead to odd elasticity \cite{scheibner2020odd,kole2024chirality,surowka2023odd,zhao2026nonlinear,lee2026elastodynamics,maire2026topology}. Unlike conventional elastic interactions, which derive from a potential energy and are therefore described by a symmetric stiffness matrix, odd elastic couplings are non-reciprocal and cannot generally be derived from a scalar potential \cite{ishimoto2022self,alexander2021layered,braverman2021topological,shankar2024active,lee2026odd,huang2025anomalous}. As a consequence, a deformation applied along one direction can generate a response in a transverse direction, and the work performed over a closed deformation cycle need not vanish. Odd elasticity therefore provides a minimal framework for describing active solids whose mechanical response intrinsically involves the injection and redistribution of energy. Even in the absence of inertia, odd elastic couplings can generate unexpected exceptional points \cite{scheibner2020odd,zhou2020non,cheng2026elastodynamics,zhou2026curved}, where two eigenvalues coalesce and change from real to complex conjugate, marking a transition from purely relaxational to oscillatory dynamics~\cite{bender1998real,heiss2012physics, fernandez2018exceptional}. 
In inertial odd-elastic solids, they can further produce dissipationless collective modes and dynamical instabilities at finite wavevectors \cite{caprini2025odd}.

Odd elastic forces, however, are not the only transverse forces that can act on a solid. A fundamentally different example is provided by gyroscopic forces, such as the Lorentz and Coriolis forces. These forces are proportional to velocity and are always orthogonal to it; consequently, they perform no work and neither inject nor dissipate energy. This is in sharp contrast with odd elastic forces, which are proportional to displacement, violate Maxwell-Betti reciprocity~\cite{maxwell1864calculation,betti1872teoria,love2013treatise}, and can perform net work over a closed deformation cycle~\cite{fruchart2021non}. 
While the effects of gyroscopic forces are well understood in passive elastic and magnetic crystals~\cite{ott2014dynamics,kahlert2012magnetizing}, and have recently been investigated in Brownian and active systems~\cite{vuijk2020lorentz,chun2018emergence,abdoli2020nondiffusive,wittmann2025confined}, their interplay with odd elasticity has, to our knowledge, not been explored.

\begin{figure*}[t]
\centering
\includegraphics[width=\textwidth]{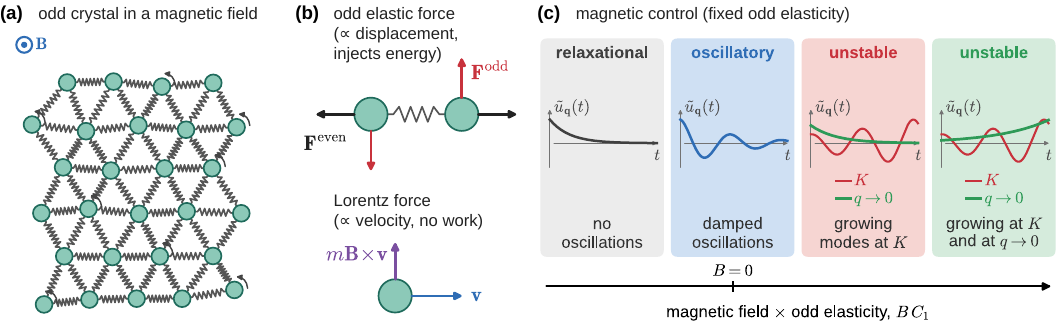} 
\caption{\textbf{Magnetic control of an odd crystal.}
(a) Triangular crystal of inertial particles connected by harmonic springs (conventional elasticity, $C_0$) and interacting through transverse odd forces (odd spring constant, $C_1$). Each particle also experiences a gyroscopic force, which can be realized either as a Lorentz force or as a Coriolis force. For notational convenience, we refer to $\mathbf B=B\hat{\mathbf z}$, perpendicular to the plane of motion, as the magnetic field and to the gyroscopic force as the Lorentz force. Curved arrows indicate the active chirality at the origin of the odd interactions. (b) Forces acting in the crystal. The central force $\mathbf F^{\rm even}$ acts along the bond, while the odd force $\mathbf F^{\rm odd}$ is perpendicular to it and proportional to the relative displacement; being non-conservative, it can inject energy into the system. The Lorentz force $m\mathbf B\times\mathbf v$ is perpendicular to the velocity and performs no work. (c) Schematic of the main results. At fixed odd elasticity, the product $BC_1$ between magnetic field and odd spring constant controls the relaxation of a perturbation $\tilde u_\bq(t)$. Increasing $BC_1$, the relaxation turns from purely relaxational to damped oscillatory beyond an exceptional point. For sufficiently large $BC_1>0$, the modes at the corners $K$ of the Brillouin zone grow (red curves), while long-wavelength perturbations, $q\to0$, still decay (green curve). For $BC_1/(\gamma C_0)>1$, also the long-wavelength perturbations grow, and the crystal is unstable both at $K$ and at $q\to0$. The sequence shown corresponds to an odd elasticity large enough for the crystal to be already oscillatory at vanishing magnetic field.}
\label{fig:overview}
\end{figure*}

We consider a two-dimensional triangular crystal of particles (Fig.~\ref{fig:overview}(a)) with underdamped Langevin dynamics, interacting
through both conventional (even) and odd elastic nearest-neighbour forces~\cite{caporusso2024phase,caprini2025odd}, and subject to a gyroscopic (Lorentz-like) force generated by a magnetic field orthogonal to the plane of motion (Fig.~\ref{fig:overview}(b)).
Although Lorentz forces cannot modify the steady-state fluctuations of passive crystals, because they do no work, they can act on the circulating currents that odd elasticity generates in the steady state.
We show that the interplay between odd elasticity and Lorentz force produces an instability at vanishing wavevector, associated with an enhancement of the long-wavelength fluctuations and a vanishing relaxation rate of the long-wavelength modes, which is absent both in ordinary magnetized crystals and in non-magnetic odd-elastic solids, where odd elasticity only destabilizes the crystal at finite wavevector. Moreover, the magnetic field provides an external handle for tuning relaxational, oscillatory and unstable phases (Fig.~\ref{fig:overview}(c)), changing the location of exceptional points and of the finite-wavevector instabilities already known to arise from odd elasticity alone, and thus the spatial velocity correlations and the structure factor of the crystal. 

The paper is organized as follows. Section~\ref{Oddelasticmodel} introduces the model, and Sec.~\ref{Dynamicintheslattice} analyzes its spectrum, identifying exceptional points and instabilities. Section~\ref{Staticproperties} discusses the steady-state properties, including the long-wavelength instability, spatial velocity correlations and the static structure factor, while Sec.~\ref{Timedependentcorrelations} addresses time-dependent correlations. Finally, Sec.~\ref{Conclusions} summarizes our results and discusses possible experimental realizations.

\section{Odd Elasticity Meets the Lorentz Force}
\label{Oddelasticmodel}

Before introducing the model, it is useful to recall the fundamental distinction between conventional and odd elasticity. In ordinary elastic materials, internal forces derive from a potential energy, implying a symmetric stiffness matrix. Consequently, the mechanical response is reciprocal and, in the absence of external dissipation, energy is conserved. By contrast, odd-elastic materials possess antisymmetric elastic couplings that cannot generally be derived from a scalar potential~\cite{fruchart2023odd}. These couplings violate Maxwell-Betti reciprocity \cite{maxwell1864calculation,betti1872teoria,love2013treatise} and allow the material to inject, extract, and redistribute energy through internal active processes\footnote{Maxwell stated that given two arbitrary points, A and B in an elastic body, the displacement of point A that results from a force applied at point B, is equal to the displacement produced at point B by the application of the same force at point A.}. As a result, the response to an applied deformation is no longer symmetric: a force acting along one direction may induce a displacement in another, even perpendicular, direction, and the influence of one degree of freedom on another need not be reciprocal. 

In this study, we consider a microscopic model for an odd active crystal in which $N=L^2$ particles are arranged on a two-dimensional triangular lattice (Appendix~\ref{app:lattice}), with positions $\bR_{\bn}$, with $\bn=(n_x, n_y)$ and $n_x, n_y=1,\dots, L$. Each particle follows an inertial dynamics for its velocity $\mathbf{v}_{\bn} = \dot{\bR}_{\bn}$, in which energy is injected through a Gaussian white noise vector, $\boldsymbol{\xi}_{\bn}(t)$, with zero average and correlations $\langle\xi^\alpha_\bn(t)\,\xi^\beta_\mb(t')\rangle=\delta_{\bn\mb}\,\delta^{\alpha\beta}\,\delta(t-t')$, and dissipated through a Stokes friction force $- m \gamma \mathbf{v}$, where $m$ denotes the particle mass and $\gamma$ is the friction rate, so that $1/\gamma$ corresponds to the inertial time.
The particle dynamics is given by an underdamped equation of motion
\begin{equation}
m \dot{\mathbf{v}}_{\bn} = \sum_{\mb\neq\bn} \Bigl( {\bf F}^{\mathrm{even}}_{\bn\mb} + {\bf F}^{\mathrm{odd}}_{\bn\mb} \Bigr) + m\,\mathbf{B}\times\mathbf{v}_{\bn} - m\gamma\,\mathbf{v}_{\bn} + \sqrt{2m\gamma T}\, \boldsymbol{\xi}_{\bn}(t) \,,
\label{ALangevin1}
\end{equation}
where $T$ is the temperature of the environment.
The term $m\,\mathbf{B}\times\mathbf{v}_n$ is a gyroscopic force, proportional and orthogonal to the particle velocity, with $\mathbf{B}=B\,\hat{\mathbf{z}}$ directed along the axis $\hat{\mathbf{z}}$ orthogonal to the plane of motion. This force admits two physical interpretations. It can be regarded as a Lorentz force generated by a uniform magnetic field if the particles are charged and their Coulomb interactions are screened, as in complex plasmas, so that the short-range part of the electrostatic interaction can be absorbed into the effective spring constants; in this case, charge and mass are absorbed in the definition of $B$, which corresponds to the cyclotron frequency. Alternatively, it can be regarded as the Coriolis force experienced by the particles in a frame rotating with angular velocity $\Omega_\mathrm{rot}$ around $\hat{\mathbf{z}}$, with $B=-2\Omega_\mathrm{rot}$, which provides one of the most natural experimental realizations. In both cases, $B$ has the dimensions of a frequency, however, in the following, we refer to $\mathbf{B}$ as the magnetic field and to $m\,\mathbf{B}\times\mathbf{v}_n$ as the Lorentz force, for simplicity of notation.

%The term $m\,\mathbf{B}\times\mathbf{v}_{\bn}$ is a gyroscopic (Lorentz-like) force, proportional to the velocity and generated by a uniform magnetic field $\mathbf{B}=B \hat{\mathbf{z}}$, directed along the axis $\hat{\mathbf{z}}$ orthogonal to the plane of motion.  Since the particle {\color{red} charge} and mass are absorbed in its definition, $B$ has the dimensions of a frequency and corresponds to the cyclotron frequency. The model is sketched in Fig.~\ref{fig:overview}(a), while Fig.~\ref{fig:overview}(b) compares the transverse forces at play.

The particles interact through both conservative central forces and non-conservative transverse forces that depend on the relative distance $|\bR_{\bn}-\bR_{\mb}|$. For simplicity, we restrict the interactions to the six nearest neighbors of the triangular lattice. Although sufficiently large displacements could in principle bring more distant particles closer than nearest neighbors, such configurations are neglected in the present analysis.
The conservative interaction generates reciprocal forces aligned with the bond direction and conserves energy~\cite{caprini2025odd}. The odd interaction produces forces perpendicular to the bond and is intrinsically non-conservative. Such forces generally violate angular momentum conservation and can perform net work on the system. A complete analytical treatment becomes possible by linearizing the interactions around the reference crystalline configuration~\cite{roberts2026exact}. Denoting by $\rbu_{\bn}$ the displacement of particle $\bn$ from its equilibrium position $\bR^0_{\bn}$, the linearized forces take the form (see Appendix~\ref{Linearization})
\begin{subequations}
\begin{align}
{\bf F}^{\mathrm{even}}_{\bn\mb} &= C_0 \Bigl[ (\rbu_{\mb}-\rbu_{\bn})\cdot {\bf e}_{\bn\mb} \Bigr] {\bf e}_{\bn\mb}, \label{forcevitelli1} \\
{\bf F}^{\mathrm{odd}}_{\bn\mb} &= C_1 \Bigl[ (\rbu_{\mb}-\rbu_{\bn})\cdot {\bf e}_{\bn\mb} \Bigr] (\hat{\bf z}\times {\bf e}_{\bn\mb}), \label{forcevitelli2}
\end{align}
\end{subequations}
where $C_0$ and $C_1$ are the even and odd spring constants, respectively, and $\hat{\bf z}$ is the unit vector normal to the plane. The unit vectors ${\bf e}_{\bn\mb}$ point from site $\mb$ to site $\bn$ in the undeformed lattice and, thus, can be expressed as:
\begin{equation}
{\bf e}_{\bn\mb} = \frac{\bR^0_{\bn}-\bR^0_{\mb}} {|\bR^0_{\bn}-\bR^0_{\mb}|} \,,
\end{equation}
with a specific form fixed by the triangular lattice.
We remark that the Lorentz force and the odd interactions are fundamentally different, even though both are transverse.
Indeed, the former is a single-body force proportional to the velocity, while the latter is a two-body interaction depending on the particle displacement (Fig.~\ref{fig:overview}(b)).
As is well known, the Lorentz force does not enter the energy balance because it performs no work. The odd force, instead, affects the kinetic energy of the system even in the steady state (see Appendix \ref{app_energetic_balance} for the derivation of this result in Fourier space), so that the kinetic temperature deviates from the temperature of the environment.

The dynamics proposed in Eq.~\eqref{ALangevin1} corresponds to a microscopic model for an odd active crystal, whose elastodynamics is characterized by conventional and odd elasticity.
Specifically, ${\bf F}^{\mathrm{even}}_{\bn\mb}$ (Eq.~\eqref{forcevitelli1}) generates conventional elasticity, fixing the bulk and shear moduli, 
while ${\bf F}^{\mathrm{odd}}_{\bn\mb}$ (Eq.~\eqref{forcevitelli2}) gives rise to odd elasticity, introducing two additional coefficients, namely the odd bulk and the odd shear moduli. The exact calculations of these coefficients in terms of the model parameters are reported in Appendix \ref{elasticmoduli}: for the present lattice, the bulk and shear moduli are $B_0=2\mu=\frac{\sqrt3}{2}C_0$, while the odd bulk and odd shear moduli are $A=2K_0=\frac{\sqrt3}{2}C_1$ (see also Ref.~\cite{scheibner2020odd}).

\subsection{Fourier space dynamics}
The linearity of the dynamics allows us to conveniently express it in Fourier space, by introducing a wave vector $\mathbf{q}$ and denoting by a tilde the spatial Fourier transform of the displacement, $\tilde{\mathbf{u}}_{\mathbf{q}}$, and of the velocity, $\tilde{\mathbf{v}}_{\mathbf{q}}=\dot{\tilde{\mathbf{u}}}_{\mathbf{q}}$ (see Appendix~\ref{app_energetic_balance}, Eq.~\eqref{eq:FT_def}, for the definition of the Fourier transform).
By introducing the Fourier transform of the noise vector $\tilde{\boldsymbol{\xi}}_\bq(t)$ such that $\langle \tilde{\xi}^{\alpha}_\bq(t) \tilde{\xi}^{\beta}_{\bq'}(t') \rangle= \delta(t-t')\,\delta^{\alpha\beta}\, \delta_{\bq+\bq',0}$, with $\alpha,\beta=x,y$,  the Langevin dynamics in matrix form, in terms of displacement and velocity variables, reads
\begin{equation}
\label{eq:fourier_space_eq}
\frac{d}{dt }\begin{pmatrix} \tilde{u}_\bq^x \\ 
\tilde{v}_\bq^x \\ \tilde{u}_\bq^y \\ \tilde{v}_\bq^y \end{pmatrix}
=-\begin{pmatrix} 0 & -1 & 0 &0 \\ -M_{xx}& \gamma & -M_{xy}&B\\ 0 & 0& 0 &-1 \\ -M_{yx} &-B & -M_{yy}& \gamma \end{pmatrix}
\begin{pmatrix} \tilde{u}_\bq^x \\ \tilde{v}_\bq^x \\ 
\tilde{u}_\bq^y \\ \tilde{v}_\bq^y \end{pmatrix}
+\sqrt{\frac{2\gamma T}{m}}
\begin{pmatrix} 0 \\ \tilde{\xi}_\bq^x \\ 0 \\ \tilde{\xi}_\bq^y \end{pmatrix},
\end{equation}
where the two-dimensional matrix $\mathbf{M}$ can be obtained from the interactions (Eqs.\eqref{forcevitelli1} and \eqref{forcevitelli2}) and corresponds to the interparticle forces, linear in the displacement. The full expression of $\mathbf{M}$ is cumbersome and is reported in Appendix~\ref{matrixelements}. Here, we focus on two limits: the small wavevector limit $\mathbf{q} \to 0$ and the large wavevector limit $\mathbf{q}\to \mathbf{K}$, where the vector $\mathbf{q}=\bK$ denotes a corner of the Brillouin zone, e.g.\ $\bK=(4\pi/3a,0)$.
It is also convenient to define the two combinations $\omega_\bq^2=-\Tr(M)/2$ and $\alpha_\bq^2=(M_\bq^{xy}-M_\bq^{yx})/2$ of the elements of $\mathbf{M}$.

In the small wavevector limit ($\mathbf{q}\to0$) and for a wavevector directed along the $x$ axis, $\bq=q\,\hat{\mathbf x}$, the diagonal elements of this matrix are not equal and are given by
\begin{equation}
M_\bq^{xx}\approx -\frac{9 C_0}{8m} q^2 \neq -\frac{3 C_0}{8m} q^2 = M_\bq^{yy} \,.
\end{equation} 
Apart from numerical prefactors, their values are fixed by the conventional spring constant $C_0$ alone. For this choice of $\bq$, $M^{xx}_\bq$ and $M^{yy}_\bq$ are the longitudinal and transverse stiffnesses, respectively. Their ratio, equal to 3, does not reflect an anisotropy of the lattice: at long wavelengths the triangular lattice is elastically isotropic, and the same longitudinal and transverse values are obtained for any direction of $\bq$.

The presence of odd interactions introduces off-diagonal elements in the matrix $\mathbf{M}$, proportional to $C_1$, which, in the small wavevector limit, take the simple form
\begin{subequations}
\begin{flalign} 
&M_{\bq}^{xy}\approx \frac{3 C_1}{8m} q^2\\
&M_{\bq}^{yx}\approx-\frac{9 C_1}{8m} q^2 \,.
\end{flalign}
\end{subequations}
These elements have opposite sign, reflecting the occurrence of odd elasticity, but different magnitudes, since they couple the longitudinal and transverse components, which have different stiffnesses.

At the corners of the Brillouin zone, $\bq=\bK$, symmetry fixes the matrix elements of $\mathbf{M}$ to
\begin{subequations}
\label{eq:expression_one}
\begin{flalign}
\omega_\bK^2 &=-M^{xx}_\bK=-M^{yy}_\bK= \frac{9}{2}\frac{C_0}{m}, \label{mk2}\\
\alpha_\bK^2 &=M^{xy}_\bK=-M^{yx}_\bK= \frac{9}{2}\frac{C_1}{m} \label{mk3}\,.
\end{flalign}
\end{subequations}
At $\bq=\bK$, the matrix $\mathbf{M}$ thus takes the isotropic form $\mathbf{M}_\bK=-\omega_\bK^2\,\mathbf{I}+\alpha_\bK^2\,\boldsymbol{\epsilon}$, with $\mathbf{I}$ the identity matrix and $\boldsymbol{\epsilon}=\bigl(\begin{smallmatrix}0&1\\-1&0\end{smallmatrix}\bigr)$ the two-dimensional Levi-Civita tensor.

%%%%%%%%%%%%%%%%%%%%%%%%%%%%%%%%%%%%
%%%%%%%%%%%%%%%%%%%%%%%%%%%%%%%%%%%%

\section{Spectrum of the Odd-Elastic Magnetized Crystal}
\label{Dynamicintheslattice}

Before turning to the odd elastic crystal, it is useful to recall the reference case. For a standard harmonic crystal, without magnetic field ($B=0$) and without odd elasticity ($C_1=0$), the spectrum contains two branches, longitudinal and transverse, associated with compression and shear waves; friction $\gamma$ turns these into diffusive rather than propagating modes at long wavelengths. 
The long-wavelength spectrum separates into two diffusive Goldstone modes with eigenvalues $\lambda\sim-q^2$ and two fast relaxational modes with $\lambda=-\gamma$, defining two well-separated time scales, $\tau_s\sim q^{-2}$ and $\tau_h\sim 1/\gamma$, with $\tau_s\gg\tau_h$ as $q\to0$.

The dynamics generated by Eq.~\eqref{ALangevin1} is intrinsically non-Hermitian, owing to (i) viscous dissipation, (ii) the non-reciprocal, odd coupling between Cartesian displacement components, and (iii) the Lorentz force; consequently, eigenmodes generically combine damping and rotation.
The central result of this Section is that the competition between ordinary dissipation, odd elasticity, and the Lorentz force reorganizes this spectrum into two qualitatively different regimes, connected by the appearance of exceptional points. At long wavelength ($\bq\to0$), the crystal crosses over from purely relaxational (overdamped) decay to damped-oscillatory decay as the relative strength of $C_1$ and $B$ is tuned; at the boundary of the Brillouin zone ($\bq\to\bK$), the same competition can drive the damping of a pair of modes exactly to zero, producing a dissipationless finite-frequency mode and, beyond that value of $C_1/C_0$, a genuine dynamical instability. Figure~\ref{fig:phasediagram}(a) summarizes both boundaries -- together with a third one, discussed in Sec.~\ref{Staticproperties}, associated with the divergence of the forward-scattering intensity -- in the plane of magnetic field and odd elasticity $(B/\gamma,\,C_1/C_0)$, and provides a map to which we refer throughout this Section.

\subsection{Long-wavelength regime: from diffusive to damped-oscillatory dynamics}
\label{Exceptionalline}

Introducing a Fourier representation of the displacement field and seeking normal-mode solutions of Eq.~\eqref{ALangevin1} leads, after eliminating the mode amplitudes, to a characteristic polynomial that is quartic in the eigenvalue $\lambda$ (Appendix~\ref{app:spectrum_derivation}, Eq.~\eqref{eqquartogrado_final}); this equation governs the spectrum at any wavevector $\bq$ and depends on $\bq$ only through the elements of the dynamical matrix $M_\bq$, whose long-wavelength expansion is likewise reported in Appendix~\ref{app:spectrum_derivation}. At $q\to0$ the quartic factorizes exactly into two vanishing (double) roots and two complex roots $\lambda^{(3,4)}=-\gamma\pm iB$: the former are the soft hydrodynamic modes of interest here, the latter fast cyclotron-like modes set by the magnetic field alone.

\vskip8pt
\noindent
\textbf{Displacement spectrum for small wavevector -- } In the overdamped regime ($\gamma\gg\omega_\bq$), adiabatic elimination of the fast pair (Appendix~\ref{app:spectrum_derivation}) reduces the quartic to a quadratic equation for the two soft eigenvalues, with solutions
\begin{equation}
\lambda_\pm= -\nu\Bigl[p\pm\sqrt{p^2-r}\Bigr],
\label{diffusiveeigenvalues}
\end{equation}
where $\nu\propto q^2$ sets the overall diffusive rate, while $p$ and $r$ are $\mathbf{q}$-independent and play the role of an effective damping rate and an effective restoring strength, respectively. They depend on friction, magnetic field, spring constants and mass through
\begin{equation}
\nu=\frac{3}{4}\frac{q^2}{\gamma^2+B^2}, \qquad
p=\gamma \frac{C_0}{m}-B \frac{C_1}{m}, \qquad
r=\frac{3}{4}(\gamma^2+B^2)\left[\Bigl(\frac{C_0}{m}\Bigr)^2+\Bigl(\frac{C_1}{m}\Bigr)^2\right].
\label{eq:nupr}
\end{equation}
Crucially, $p$ combines the ordinary friction $\gamma C_0/m$ with a magnetic contribution $-BC_1/m$ that exists only because odd elasticity generates a circulating current for the Lorentz force to act on; without $C_1$, $B$ would not appear in $p$ at all.

\vskip8pt
\noindent
\textbf{Exceptional points -- } 
This competition has a sharp physical consequence. When $p^2>r$, the two roots in Eq.~\eqref{diffusiveeigenvalues} are real and negative: perturbations relax monotonically, as in an ordinary overdamped solid. When $p^2<r$, they form a complex-conjugate pair and relaxation turns into damped oscillation. The boundary $p^2=r$ defines a curve of \emph{exceptional points} in the plane of magnetic field and ratio between odd and even spring constants $(B/\gamma,\,C_1/C_0)$, where the two eigenvalues coalesce and the effective $2\times2$ evolution operator of the soft modes becomes non-diagonalizable:
\begin{equation}
\frac{C_1}{C_0}
=
-\frac{1}{3 - \frac{B^2}{\gamma^2}}
\left(
4\frac{B}{\gamma}
\pm \sqrt{3}\left(1+\frac{B^2}{\gamma^2}\right)
\right).
\label{exceptionalline}
\end{equation}
At $B=0$ this reduces to the known non-magnetic result, $C_1/C_0=\pm1/\sqrt3$. Equation~\eqref{exceptionalline} shows that the magnetic field is not a passive spectator: a field aligned with the odd-elastic chirality ($BC_1>0$) lowers the odd-elastic strength needed to reach the exceptional point, so that oscillations set in already for weaker $C_1$; a field of opposite sign has the reverse effect, delaying the onset of oscillatory relaxation. For $|B|>\gamma/\sqrt3$, the long-wavelength modes are oscillatory even in the absence of odd elasticity, since the Lorentz force alone mixes longitudinal and transverse modes, which have different stiffnesses. In this case, odd elasticity of opposite sign opens a window of purely relaxational dynamics, e.g.\ $2-\sqrt3<C_1/C_0<2+\sqrt3$ for $B/\gamma=-1$ (see Fig.~\ref{fig:phasediagram}(b)). The solid blue curve in Fig.~\ref{fig:phasediagram}(a) traces this boundary, separating the diffusive (grey) from the damped-oscillatory (light blue) regime. This is the first manifestation of a mechanism that recurs throughout the paper: the Lorentz force, unable by itself to alter the elastic response of a passive crystal, becomes an active player once odd elasticity supplies it with a current to act on. 

\subsection{Finite-wavevector regime: non-dissipative modes and the Brillouin-zone instability}
\label{sec:Kpoint_instability}

At the boundary of the Brillouin zone the separation of time scales invoked above no longer holds: inertia becomes essential, and the full quartic equation (Appendix~\ref{app:spectrum_derivation}, Eq.~\eqref{eqquartogrado_final}) must be solved without approximation. At the six corners $\bq=\bK$, lattice symmetry constrains the dynamical matrix to the simple isotropic form of Eqs.~\eqref{mk2}--\eqref{mk3}, and the quartic factorizes exactly into two complex-conjugate quadratics (Appendix~\ref{Kpoint}). The real parts of the two resulting mode pairs sum to $-\gamma$, as fixed by the trace of the drift matrix. One pair has ${\rm Re}\,\lambda=-\gamma$, while other becomes \emph{undamped}, with ${\rm Re}\,\lambda=0$, when
\begin{equation}
(\alpha_\bK^2)^2 + \gamma B \alpha_\bK^2 = \gamma^2 \omega_\bK^2.
\label{Kinstability0}
\end{equation}
This condition has a direct physical reading: $\gamma^2\omega_\bK^2$ is the ordinary dissipation acting on the ordinary elastic restoring force, while $(\alpha_\bK^2)^2$ is the destabilizing contribution of odd elasticity alone, and $\gamma B\alpha_\bK^2$ is the effect of the magnetic field acting on the current generated by odd elasticity, which can either reinforce or oppose it depending on the sign of $BC_1$. When the two balance exactly, a pair of modes that would ordinarily relax stops dissipating altogether and oscillates forever at ${\rm Im}\,\lambda=\pm\tfrac12(B-\sqrt{B^2+4\omega_\bK^2})$ -- a genuine finite-wavevector analogue of the $\bq=0$ Goldstone mode. Beyond this point, the real part of the eigenvalue ${\rm Re}\,\lambda$ turns positive: linear theory predicts a dynamical instability at $\bq\simeq\bK$ that can be arrested only by nonlinear effects. This is qualitatively different from an ordinary Goldstone mode: the suppression of dissipation occurs at \emph{finite} frequency and \emph{finite} wavevector, a hallmark of the underdamped dynamics considered here, and when it occurs the intensity of the corresponding wave grows while its spectral width narrows.

By using the definitions \eqref{eq:expression_one}, the finite-wavevector instability condition~\eqref{Kinstability0} can be recast as a condition on the ratio between odd and even spring constants, $C_1/C_0$:
\begin{equation}
\frac{C_1}{C_0}=-\frac{\frac{B}{\gamma} \pm \sqrt{(B/\gamma)^2+ 2 w}}{w}, \qquad w=9 \frac{C_0}{m \gamma^2},
\label{eq:Kinstability_ratio}
\end{equation}
shown as the dashed red curve in Fig.~\ref{fig:phasediagram}(a). A magnetic field aligned with the odd-elastic chirality pushes the onset of this instability towards ever smaller $|C_1/C_0|$, so that for $B\to\infty$ an arbitrarily weak odd elasticity suffices to destabilize the crystal at the zone boundary; a field of opposite sign instead stabilizes the solid, moving the instability to larger $|C_1/C_0|$. As shown in Sec.~\ref{Staticproperties}, the equal-time correlations computed from the same dynamics diverge exactly on this line: a vanishing real part of $\lambda$ means that continuous thermal-noise injection is no longer balanced by dissipation.
 
The exceptional-point line, Eq.~\eqref{exceptionalline}, and the finite-wavevector instability line, Eq.~\eqref{eq:Kinstability_ratio}, are collected in the phase diagram of Fig.~\ref{fig:phasediagram}(a), discussed below, together with a third boundary, $BC_1/(\gamma C_0)=1$, which marks the instability of the long-wavelength modes. Its derivation, and its different nature compared with the instability at $\bK$, are discussed in Sec.~\ref{Staticproperties}.
Sections~\ref{Staticproperties} and~\ref{Timedependentcorrelations} translate these spectral boundaries into observable signatures in the static and dynamic correlation functions.

\begin{figure*}[t]
\centering
\includegraphics[width=\textwidth]{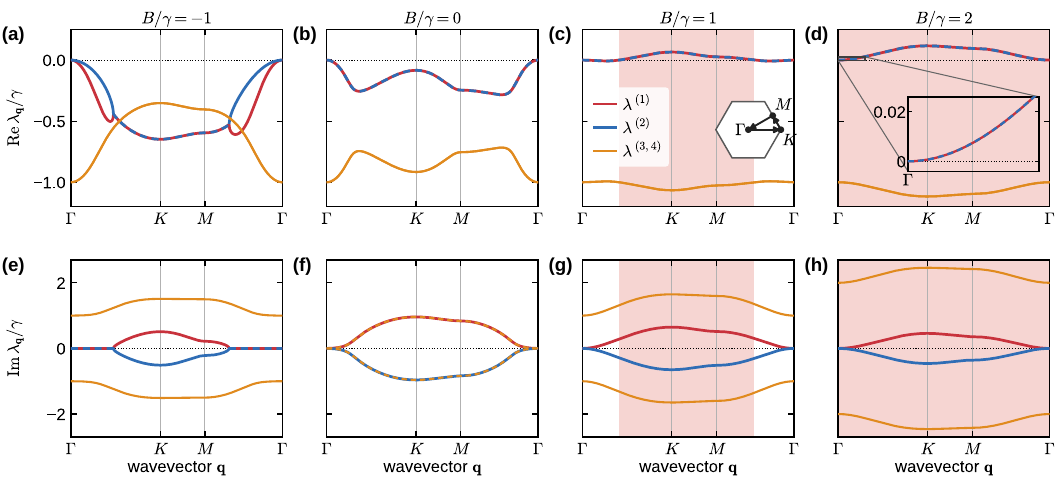}
\caption{\textbf{Spectrum of the odd crystal in a magnetic field.}
Real (a)--(d) and imaginary (e)--(h) parts of the four eigenvalues $\lambda_\bq$ of the linearized
dynamics, i.e.\ the roots of Eq.~\eqref{eqquartogrado_final} (see Appendix~\ref{app:spectrum_derivation}). The eigenvalues are shown along the high-symmetry path
$\Gamma\to K\to M\to\Gamma$ of the first Brillouin zone (inset of panel (c)), with
$\Gamma=(0,0)$, $K=(4\pi/3a,0)$ and $M=(\pi/a,\pi/\sqrt3a)$, and the horizontal axis is the arc length
along the path. Red and blue: acoustic branches $\lambda^{(1)}_\bq$, $\lambda^{(2)}_\bq$, vanishing at
$\Gamma$; orange: cyclotron-like branches $\lambda^{(3,4)}_\bq$, equal to $-\gamma\pm iB$ at $\Gamma$. 
The legend shown in (c) applies to all panels.
Dashed lines mark branches that coincide with another branch (complex-conjugate pairs have equal
real parts). Shaded areas: wavevectors where the crystal is linearly unstable, ${\rm Re}\,\lambda_\bq>0$. The inset in (d) magnifies the region close to $\Gamma$ marked by the small rectangle, where ${\rm Re}\,\lambda_\bq>0$ down to $q\to0$.
Parameters: $C_1/C_0=0.8$, $w=9C_0/(m\gamma^2)=2$, and $B/\gamma=-1$ (a,e), $0$ (b,f), $1$ (c,g), $2$ (d,h),
corresponding to the white symbols in Fig.~\ref{fig:phasediagram}(a).}
\label{fig:spectrum}
\end{figure*}

\vskip8pt
\noindent
\textbf{Full spectrum --} The two analytical cases discussed above are two limits of the full spectrum of the linearized dynamics, shown in Fig.~\ref{fig:spectrum}. There, the four eigenvalues $\lambda_\bq$, i.e.\ the roots of Eq.~\eqref{eqquartogrado_final}, are obtained numerically along the high-symmetry path $\Gamma\to K\to M\to\Gamma$ of the first Brillouin zone (inset), with $\Gamma=(0,0)$, $K=(4\pi/3a,0)$ and $M=(\pi/a,\pi/\sqrt3 a)$, for fixed odd elasticity, $C_1/C_0$, and four different values of the magnetic field, $B/\gamma$ (symbols in Fig.~\ref{fig:phasediagram}(a)). Independently of the magnetic field, two branches, $\lambda^{(1)}_\bq$ and $\lambda^{(2)}_\bq$, vanish at $\Gamma$ and correspond to the diffusive (Goldstone) modes of Eq.~\eqref{diffusiveeigenvalues}, while the other two, $\lambda^{(3,4)}_\bq$, start from the cyclotron-like values $-\gamma\pm iB$. Since the drift matrix is real, eigenvalues are either real or come in complex-conjugate pairs, with equal real parts and opposite imaginary parts. 

We start by discussing an odd crystal in the absence of magnetic field (Figs.~\ref{fig:spectrum}(b) and (f)), as a reference case. Since $C_1/C_0>1/\sqrt3$, the acoustic pair is complex already at small $q$: perturbations relax through damped oscillations at every wavevector, and the least damped mode, at $K$, has ${\rm Re}\,\lambda_\bK <0$. Increasing the odd elasticity further, this mode becomes undamped at $C_1/C_0=1$, the instability threshold at $B=0$. However, the mode at $\Gamma$ remains always stable as predicted in Ref.\cite{caprini2025odd}.

The behavior of the spectrum depends on the sign of magnetic field compared to $C_1$, which is chosen positive as a reference.
For negative magnetic field ($B/\gamma=-1$, Figs.~\ref{fig:spectrum}(a) and (e)), a field opposite to the odd chirality, the two acoustic eigenvalues are real and distinct close to $\Gamma$: long-wavelength perturbations relax without oscillating. At finite wavevector they coalesce at an exceptional point and then form a complex-conjugate pair, whose frequencies open up symmetrically around zero in panel (e). All real parts remain negative, and close to $K$ the least damped mode is the cyclotron-like one: by bending the particle trajectories in the sense opposite to the odd forces, the field partly compensates their effect. 

A field aligned with the odd chirality, ($B/\gamma=1$, Figs.~\ref{fig:spectrum}(c) and (g)), pushes the real part of the acoustic pair above zero in a broad region around $K$ and $M$ (shaded): the crystal is linearly unstable, since ${\rm Re}\,\lambda_\bK>0$ at the zone corner, in agreement with Eq.~\eqref{Kinstability0}. Intuitively, the field now bends the trajectories in the same sense as the odd forces, which can thus inject energy more efficiently than friction removes it. The cyclotron branch, instead, becomes more strongly damped. For an even stronger field, $B/\gamma=2$ (Figs.~\ref{fig:spectrum}(d) and (h)), the product $BC_1/(\gamma C_0)=1.6$ exceeds one, and the real part of the acoustic pair is positive along the whole path, including the neighbourhood of $\Gamma$, where it vanishes as $q^2$ (inset of Fig.~\ref{fig:spectrum}(d)). In this case, the crystal is unstable not only around the zone corners but also at long wavelengths, as expected from the condition $BC_1/(\gamma C_0)>1$ derived in Sec.~\ref{Staticproperties}. The magnetic field alone thus drives the spectrum from relaxational, to oscillatory, to unstable at finite wavevector, as anticipated in Fig.~\ref{fig:overview}(c), and, for $BC_1/(\gamma C_0)>1$, unstable also at $q\to0$.

\begin{figure*}[t]
\centering
\includegraphics[width=\textwidth]{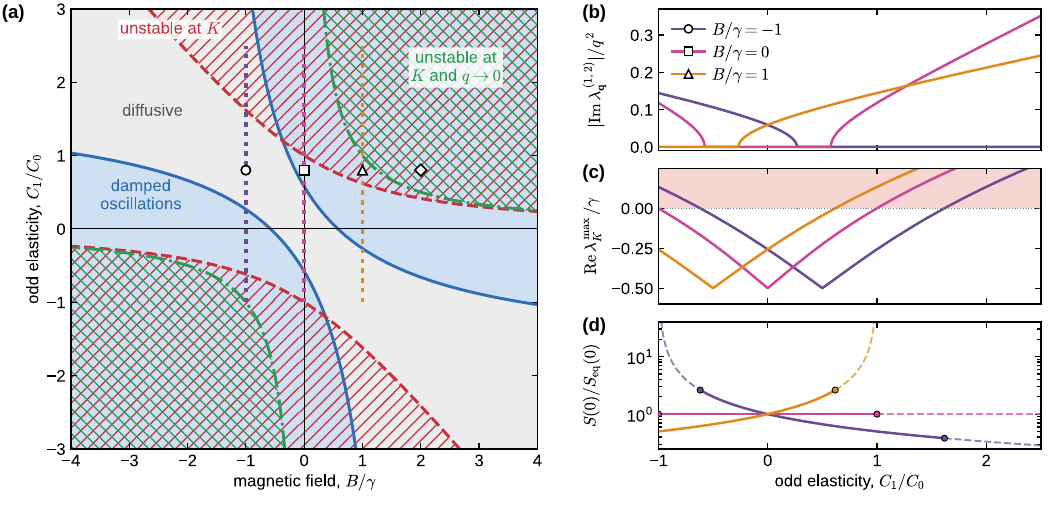}
\caption{\textbf{Phase diagram.}
(a) Phase diagram of the odd crystal as a function of magnetic field, $B/\gamma$, and odd elasticity, $C_1/C_0$. The exceptional-point line (solid blue), Eq.~\eqref{exceptionalline}, separates the diffusive (grey) from the damped-oscillatory (light blue) long-wavelength regime. The instability line at the Brillouin-zone corner (dashed red), Eq.~\eqref{eq:Kinstability_ratio}, bounds the region that is linearly unstable at $\bq\simeq\bK$ (red hatching). The line $BC_1/(\gamma C_0)=1$ (dash-dotted green), where the forward-scattering intensity of Eq.~\eqref{eq:S0} diverges, bounds the region in which also the long-wavelength modes are unstable (green hatching); this region lies entirely inside the region unstable at $\bK$. White symbols mark the parameters of Fig.~\ref{fig:spectrum}, and dotted vertical lines the cuts shown in (b)--(d).
(b) Frequency of the long-wavelength acoustic modes, $|{\rm Im}\,\lambda^{(1,2)}_\bq|/q^2$, obtained from the roots of Eq.~\eqref{eqquartogrado_final} (see Appendix~\ref{app:spectrum_derivation}), which vanishes in the diffusive regime.
(c) Largest real part of the eigenvalues at $\bK$, Eqs.~\eqref{undampedmode}--\eqref{dampedmode} (see Appendix~\ref{Kpoint}), with the unstable range shaded.
(d) Forward-scattering intensity normalized by its equilibrium value, $S(0)/S_{\rm eq}(0)$, Eq.~\eqref{eq:S0}. Solid lines correspond to a linearly stable crystal, dashed lines to parameters beyond the onset of the instability at $\bK$, which is marked by dots and given by Eq.~\eqref{eq:Kinstability_ratio}. Panels (b)--(d) share the legend shown in (b), whose symbols correspond to those in (a). The remaining parameters are: $w=9C_0/(m\gamma^2)=2$ in all panels, and $qa=0.02$ in (b).}
\label{fig:phasediagram}
\end{figure*}

\vskip8pt
\noindent
\textbf{Phase diagram --} Figure~\ref{fig:phasediagram}(a) summarizes how odd elasticity and magnetic field shape the dynamics of the crystal, for a representative value of $w$ (see Eq.~\eqref{eq:Kinstability_ratio}). A first observation is that the diagram is symmetric under inversion through the origin: a simultaneous reversal of both chiralities, $(B,C_1)\to(-B,-C_1)$, is equivalent to a mirror reflection of the crystal and leaves the spectrum unchanged. What matters is therefore not the sign of the field alone, but whether the field bends the trajectories in the same sense as the odd forces, $BC_1>0$, or in the opposite one, $BC_1<0$.

Along the vertical axis, $B=0$, odd elasticity alone controls the character of the relaxation. Weak odd forces only slightly couple longitudinal and transverse deformations, which relax independently and without oscillations (grey region). When this coupling overcomes the difference between the longitudinal and transverse relaxation rates, at the exceptional point $|C_1/C_0|=1/\sqrt3$, the two modes hybridize into a pair that rotates while relaxing, and perturbations decay through damped oscillations (light blue region). For even stronger odd forces, the energy injected at the Brillouin-zone corners exceeds the energy dissipated by friction, and the crystal becomes unstable at $\bK$ (red hatching). Inertia is essential for this last step: the threshold, $|C_1/C_0|=\sqrt{2/w}$, diverges in the overdamped limit $w\to0$.

The magnetic field acts in a different way. Since it performs no work, it cannot destabilize the crystal by itself, but it bends the particle trajectories. Along the horizontal axis, $C_1=0$, a sufficiently strong field turns the relaxation into damped, cyclotron-like oscillations, without affecting the stability of the crystal. In the presence of odd elasticity, instead, the field redirects the motion generated by the odd forces. For $BC_1>0$, this redirection cooperates with the odd forces: the oscillatory region extends to weaker odd elasticity, and the instability at $\bK$ moves to smaller $|C_1/C_0|$, approaching the horizontal axis for strong fields. For $BC_1<0$, the field counteracts the odd forces: the instability is delayed, and a window of purely relaxational dynamics opens at intermediate odd elasticity. Finally, for $BC_1/(\gamma C_0)>1$ (green hatching), the long-wavelength modes are unstable as well. This region lies entirely inside the region unstable at $\bK$ and, unlike the other boundaries, exists only when magnetic field and odd elasticity act together.

The side panels follow these trends along three cuts at fixed field, $B/\gamma=-1,0,1$ (dotted lines). The frequency of the long-wavelength acoustic modes, $|{\rm Im}\,\lambda^{(1,2)}_\bq|/q^2$ (Fig.~\ref{fig:phasediagram}(b)), vanishes in the relaxational windows and grows as a square root beyond the exceptional points, Eq.~\eqref{exceptionalline}. The largest real part of the eigenvalues at $\bK$ (Fig.~\ref{fig:phasediagram}(c)) crosses zero at weaker odd elasticity when the field is aligned with the odd chirality, and at stronger odd elasticity when it is opposite, in agreement with Eq.~\eqref{eq:Kinstability_ratio}; its cusps mark the exchange of the least damped pair between the acoustic and the cyclotron-like branches. The forward-scattering intensity $S(0)$, Eq.~\eqref{eq:S0} (Fig.~\ref{fig:phasediagram}(d)), is not affected by odd elasticity at $B=0$, increases (decreases) for $BC_1>0$ ($BC_1<0$), and diverges at $BC_1/(\gamma C_0)=1$, where the long-wavelength modes become unstable. Beyond the onset of the instability at $\bK$ (dots), the linear theory admits no steady state, and $S(0)$ is only formal (dashed lines).

%%%%%%%%%%%%%%%%%%%%%%%%%%%%%%%%%%%%%%%%%%%%
%%%%%%%%%%%%%%%%%%%%%%%%%%%%%%%%%%%%%%%%%%%%
\section{Non-Equilibrium Steady-State Properties}
\label{Staticproperties}

Section~\ref{Dynamicintheslattice} identified two boundaries in the $(B/\gamma,\,C_1/C_0)$ plane -- the exceptional-point line and the Brillouin-zone instability -- from the spectrum of the linearized dynamics alone. We now show that the Brillouin-zone instability reappears as a divergence of the steady-state fluctuations, and that a third, purely long-wavelength boundary emerges, as anticipated in Fig.~\ref{fig:phasediagram}(a). Equal-time correlations can be obtained by solving the Lyapunov equation for the steady-state covariance of Eq.~\eqref{eq:fourier_space_eq} (Appendix~\ref{Lyapunovequations}). Since the resulting expressions are algebraically cumbersome for generic anisotropic $\bq$ (because $M_\bq^{xx}\neq M_\bq^{yy}$ and $M_\bq^{xy}\neq-M_\bq^{yx}$ away from the special limits already met in Sec.~\ref{Dynamicintheslattice}), we again restrict the discussion to $\bq\to0$ and $\bq\simeq\bK$.

\subsection{Long-wavelength regime: displacement fluctuations and the compressibility instability}

\noindent
\textbf{Expression of the displacement correlations in the small wavevector regime --}
In the long-wavelength regime for $\gamma\gg\omega_\bq$, which applies to the small-wavevector modes, $\bq\to0$, the velocity degrees of freedom can be adiabatically eliminated (Appendix~\ref{Lyapunovequations}). This procedure reduces the problem to a $2\times2$ Lyapunov equation for the displacement correlations that can be conveniently projected onto the longitudinal and transverse modes in the Fourier space. Longitudinal and transverse components of the displacement correlations are defined as $\Sigma_{LL}(\mathbf{q}) = \sum_{\alpha,\beta}\hat{q}^{\alpha}\hat{q}^{\beta}\langle \tilde{u}_{\mathbf q}^{\alpha} \tilde{u}_{-\mathbf q}^{\beta}\rangle$ and $\Sigma_{TT}(\mathbf{q}) = \sum_{\alpha, \beta} (\delta^{\alpha\beta}-\hat{q}^{\alpha} \hat{q}^{\beta})\langle \tilde{u}_{\mathbf q}^{\alpha} \tilde{u}_{-\mathbf q}^{\beta}\rangle$, respectively. Here, $\hat{q}^{\alpha}=q^{\alpha}/q$ with $\alpha=x,y$, denotes the Cartesian components of the unit vector along $\mathbf q$.

The choice of these directions is particularly convenient because it allows us to find expressions for the displacement correlations that depend only on the wavevector modulus $q=|\mathbf{q}|$.
The explicit expressions for $\Sigma_{LL}(\mathbf{q})$ and $\Sigma_{TT}(\mathbf{q})$ in the small wave-vector limit are derived in Appendix~\ref{Lyapunovequations} and read
\begin{subequations}
\label{eq:long_trans_correlations}
\begin{flalign}
\Sigma_{LL}(\bq) &= \frac{8}{9}\, \frac{T}{C_0}\, \frac{1}{q^2}\, \frac{1}{1 - \frac{B}{\gamma}\frac{C_1}{C_0}}, \label{longitudinalcorrelation} \\
\Sigma_{TT}(\bq) &= \frac{8}{3}\, \frac{T}{C_0}\, \frac{1}{q^2}\, \frac{1}{1 - \frac{B}{\gamma}\frac{C_1}{C_0}}. \label{transversecorrelation}
\end{flalign}
\end{subequations}
Expressions \eqref{eq:long_trans_correlations} diverge as $1/q^2$ for $q\to0$ as in the case of passive crystals. Neither odd elasticity nor gyroscopic forces change this property and, therefore, they cannot change the divergence set by the Mermin-Wagner theorem.
The strength of $\Sigma_{LL}(\bq)$ and $\Sigma_{TT}(\bq)$ is set by thermal temperature and conventional elasticity $T/C_0$ as in passive crystals with or without magnetic field $B$ or in odd active crystals for $B=0$, and, in addition satisfies the relation $\Sigma_{TT}=3\Sigma_{LL}$. 

\vskip8pt
\noindent
\textbf{Renormalization of transverse and longitudinal displacement correlations --}
The magnetic field alone does not modify static fluctuations, but in combination with odd elasticity ($C_1\neq0$) it is able to renormalize both longitudinal and transverse correlations through the same factor $\bigl(1-\tfrac{B}{\gamma}\tfrac{C_1}{C_0}\bigr)^{-1}$. 
Fluctuations are enhanced for $BC_1>0$ and suppressed for $BC_1<0$. Formally, Eqs.~\eqref{eq:long_trans_correlations} coincide with those of a passive crystal with the reduced stiffness $C_0\bigl(1-\frac{B}{\gamma}\frac{C_1}{C_0}\bigr)$ or, equivalently, with those of a passive crystal at the effective temperature
\begin{equation}
T_{\rm eff}=\frac{T}{1-\frac{B}{\gamma}\frac{C_1}{C_0}}\,,
\label{eq:Teff}
\end{equation}
which applies to the long-wavelength displacement modes. The second interpretation is the appropriate one. Indeed, the static response of the crystal to a constant external force is not affected by the magnetic field: in a stationary deformed configuration all velocities vanish, so that the Lorentz force does not contribute, and the displacement is fixed by the dynamical matrix $\mathbf M_\bq$ alone. The elastic moduli of the crystal, Eq.~\eqref{eq:moduli_mapping}, are therefore unchanged, and the factor $\bigl(1-\frac{B}{\gamma}\frac{C_1}{C_0}\bigr)^{-1}$ does not signal a change of rigidity. It measures, instead, the violation of the equilibrium relation between equal-time fluctuations and static response, which holds for $B=0$ or $C_1=0$. Note also that $T_{\rm eff}$ only concerns the displacement fluctuations: at small wavevector, velocity fluctuations remain close to equipartition (Sec.~\ref{sec:velocitycorrelations}).
This renormalization is not an inertial effect: it depends on $B$ and $\gamma$ only through $B/\gamma$, the ratio between Lorentz and friction forces, which does not depend on the particle mass at fixed friction coefficient $m\gamma$, and it is independent of the inertial parameter $w$ introduced in Sec.~\ref{sec:Kpoint_instability}. It disappears only when friction dominates over the Lorentz force, $B/\gamma\to0$ or $C_1\to 0$.
We remark that, when $B=0$, the displacement fluctuations do not depend on the odd parameters at all, even though the \emph{dynamic} correlations do (see Sec.~\ref{Timedependentcorrelations} and the discussion reported by Ghimenti et al.~\cite{ghimenti2023sampling} for odd crystals in the overdamped regime). 

\vskip8pt
\noindent
{\textbf{Compressibility instability, at $\mathbf{q}\to0$ --}} The interplay between gyroscopic forces and odd elasticity generates a new phenomenon, absent both in passive solids with magnetic fields and in odd active crystals in the absence of these magnetic forces. When a sufficiently strong magnetic field, with the same sign as the odd elasticity, $BC_1>0$, is applied, the crystal can be unstable at any wavevector, thereby also at vanishing wavevector $\mathbf{q} \to 0$. This implies a genuinely long-wavelength, dynamical instability of the compressibility type. 
This instability is evidenced by the divergence of both longitudinal and transversal correlations when the product between magnetic field and odd elasticity is sufficiently large compared to conventional elasticity, specifically, when the following instability condition is satisfied:
\begin{equation}
\frac{B}{\gamma}\frac{C_1}{C_0}=1 \,.
\label{eq:instability_zeroq}
\end{equation}
Beyond this point, for $\tfrac{B}{\gamma}\tfrac{C_1}{C_0}>1$, the expression for $\Sigma_{LL}(\bq)$ and $\Sigma_{TT}(\bq)$ is no longer valid and the system is truly unstable also for $\mathbf{q} \to 0$.
The instability line given by Eq. \eqref{eq:instability_zeroq} has been reported in Fig.~\ref{fig:phasediagram}(a) (green curve): it requires neither the dynamics to turn oscillatory nor the spectrum to touch the imaginary axis at finite $q$, only the relaxation rate of the long-wavelength modes, proportional to $p=\gamma C_0/m-BC_1/m$ in Eq.~\eqref{diffusiveeigenvalues}, to vanish.
Since the condition depends on the product $BC_1$, this instability can be induced by a large magnetic field even in a weakly odd crystal.

\vskip8pt
\noindent
\textbf{Two instabilities of different nature --} The long-wavelength instability~\eqref{eq:instability_zeroq} is qualitatively different from the instability at the Brillouin-zone corner discussed in Sec.~\ref{sec:Kpoint_instability}. The latter is a finite-wavevector oscillatory (Hopf-like) instability, with frequency $|{\rm Im}\,\lambda_\bK|\sim\gamma$, which selects a length scale set by the lattice (the $\sqrt3\times\sqrt3$ pattern of Sec.~\ref{sec:velocitycorrelations}) and is already induced by odd elasticity alone, for $B=0$ and $|C_1/C_0|>\sqrt{2/w}$. The instability~\eqref{eq:instability_zeroq}, instead, has no selected length scale: for $BC_1/(\gamma C_0)>1$ the coefficient $p$ in Eq.~\eqref{diffusiveeigenvalues} becomes negative, and all the long-wavelength acoustic modes grow, at a positive rate ${\rm Re}\,\lambda_\pm=-\nu p\propto q^2$. It is determined by the continuum elastic moduli, $C_1/C_0$, and $B/\gamma$ only, and it requires the simultaneous presence of magnetic field and odd elasticity. Since the static response is not affected by the magnetic field, this instability is not a loss of rigidity: the undeformed crystal remains a stationary solution with finite static response, but it becomes dynamically unstable, because the Lorentz force, acting on the motion induced by the odd forces, amplifies the long-wavelength deformations instead of damping them.

The two instability lines are related by inertia. Since $BC_1/(\gamma C_0)>1$ implies $\frac{w}{2}(C_1/C_0)^2+\frac{B}{\gamma}\frac{C_1}{C_0}>1$, the region where the long-wavelength modes are unstable lies inside the region that is unstable at $\bK$, Eq.~\eqref{eq:Kinstability_ratio}. Increasing $BC_1$ from a stable crystal, the finite-wavevector instability is therefore reached first, and the line~\eqref{eq:instability_zeroq} separates two unstable regimes: a finite band of unstable modes around $\bK$, and a regime in which also the whole long-wavelength part of the spectrum is unstable. In the overdamped limit, $w=9C_0/(m\gamma^2)\to0$, the instability line at $\bK$ tends to $BC_1/(\gamma C_0)=1$, so that the two instabilities set in together, while for finite $w$ inertia anticipates the finite-wavevector instability. The fate of the crystal beyond these lines requires a nonlinear analysis, which is beyond the scope of this work; the long-wavelength instability is most directly accessible in strongly damped systems, where the two lines approach each other.

\begin{figure*}[t]
\centering
\includegraphics[width=\textwidth]{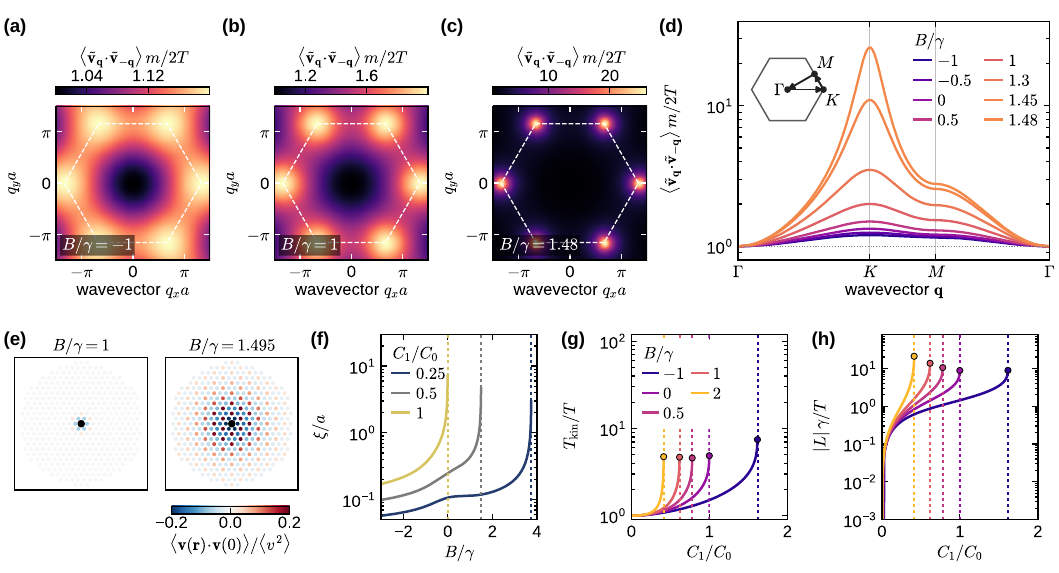}
\caption{\textbf{Spatial velocity correlations.}
(a)--(c) Velocity spectrum $\langle\tilde{\mathbf v}_\bq\cdot\tilde{\mathbf v}_{-\bq}\rangle\,m/2T$ (equal to 1 in
equilibrium) in reciprocal space for $B/\gamma=-1$ (a), $1$ (b) and $1.48$ (c), obtained from Eq.~\eqref{eq:lyap4x4} (see Appendix~\ref{Lyapunovequations}); the dashed hexagon is
the first Brillouin zone. (d) Velocity spectrum along the path $\Gamma\to K\to M\to\Gamma$ (inset) for
several $B/\gamma$. (e) Real-space velocity correlations
$\langle\mathbf v(\mathbf r)\cdot\mathbf v(0)\rangle/\langle v^2\rangle$ on the lattice sites $\mathbf r$
(black dot: $\mathbf r=0$) for $B/\gamma=1$ and $1.495$, Eq.~\eqref{eq:realspace} (see Appendix~\ref{Lyapunovequations}). (f) Velocity correlation length $\xi$, obtained from
the width of the peak at $\bK$, Eq.~\eqref{eq:xi} (see Appendix~\ref{Lyapunovequations}), as a function of $B/\gamma$ for several $C_1/C_0$. (g) Kinetic temperature $T_{\rm kin}/T$ and (h) modulus of the
angular momentum per particle, $|L|\gamma/T$, as a function of $C_1/C_0$ for several $B/\gamma$, obtained from the single-particle averages of Eq.~\eqref{eq:realspace} at $\mathbf r=0$ (see Appendix~\ref{Lyapunovequations}). 
Panels (g) and (h) share the legend shown in (g). In (f)--(h), vertical
dotted lines mark the onset of the instability at $K$, Eq.~\eqref{eq:Kinstability_ratio}, where $\xi$, $T_{\rm kin}$ and $|L|$ diverge; in (g) and (h)
the curves are computed up to $0.999$ times the critical value of $C_1/C_0$ (dots). All quantities are obtained from the exact solution of the Lyapunov equation, Eq.~\eqref{eq:lyap4x4} (Appendix~\ref{Lyapunovequations}). 
The remaining parameters are: $w=9C_0/(m\gamma^2)=2$ in all panels; $C_1/C_0=0.5$ in (a)--(e), for which the instability at $K$ occurs at $B/\gamma=1.5$. The real-space correlations in (e) are obtained by summing the exact contributions of all the wavevectors of a periodic crystal of $120\times120$ particles, and $T_{\rm kin}$ and $|L|$ for a crystal of $48\times48$ particles.}
\label{fig:velocity}
\end{figure*}

\subsection{Finite-wavevector regime: fluctuation enhancement near the Brillouin-zone boundary}

In ordinary harmonic solids, equipartition implies that displacement fluctuations decrease as $T/(m \omega_\bq^2)$ with increasing $|\bq|$, while velocity fluctuations are $\bq$-independent. Odd-elastic solids instead show strong wavevector-dependent fluctuations, enhanced near the Brillouin-zone boundary; here inertial terms cannot be neglected and the full underdamped dynamics must be used.
As shown in Appendix~\ref{Lyapunovequations}, near $\bq\simeq\bK$ the equal-time correlations take the closed form
\begin{subequations}
\begin{align}
\langle \tilde{v}_\bq^x \tilde{v}_{-\bq}^x + \tilde{v}_\bq^y \tilde{v}_{-\bq}^y \rangle &= \frac{2T}{m} \frac{\gamma^2 \omega_\bq^2 - B\gamma \alpha_\bq^2} {\gamma^2 \omega_\bq^2 - B\gamma \alpha_\bq^2 - (\alpha_\bq^2)^2}, \label{Eq:variance1} \\
\langle \tilde{u}_\bq^x \tilde{u}_{-\bq}^x + \tilde{u}_\bq^y \tilde{u}_{-\bq}^y \rangle &= \frac{2T}{m} \frac{\gamma^2} {\gamma^2 \omega_\bq^2 - B\gamma \alpha_\bq^2 - (\alpha_\bq^2)^2}, \label{Eq:variance2} \\
\langle \tilde{u}_\bq^x \tilde{v}_{-\bq}^y - \tilde{u}_\bq^y \tilde{v}_{-\bq}^x \rangle &= -\frac{2T}{m} \frac{\gamma \alpha_\bq^2} {\gamma^2 \omega_\bq^2 - B\gamma \alpha_\bq^2 - (\alpha_\bq^2)^2}. \label{Eq:variance3}
\end{align}
\end{subequations}
These expressions are exact at $\bq=\bK$ and remain quantitatively accurate in its neighborhood. Their common denominator is precisely the difference between the two sides of the finite-wavevector instability condition, Eq.~\eqref{Kinstability0}: it vanishes exactly on the dashed red curve of Fig.~\ref{fig:phasediagram}(a), and approaching that curve from the stable side the fluctuations diverge, exactly as the eigenvalue analysis of Sec.~\ref{sec:Kpoint_instability} anticipates for a mode whose damping vanishes. For odd oscillators the fluctuations remain Gaussian, but equipartition does not hold; velocities are correlated in space, as often observed in active matter, and these correlations increase with the strength of the transverse force.

In the absence of odd elasticity ($\alpha_\bq^2=0$), the velocity correlations reduce to a constant, recovering equilibrium-like equipartition; the additional $-(\alpha_\bq^2)^2$ term is a purely non-equilibrium contribution responsible for the enhancement of fluctuations near the zone boundary. This has a direct physical origin: the Lorentz force is orthogonal to velocity and cannot by itself change the energy stored in elastic deformations, so if $C_1=0$ there is no circulating current for it to act on and the equal-time covariance is independent of $B$; odd elasticity, by contrast, generates a circulating current even for $B=0$, and once such a current exists, the magnetic field acts on it and modifies the steady-state covariance.

\subsection{Spatial velocity correlations, kinetic temperature and angular momentum}
\label{sec:velocitycorrelations}

The closed-form expressions above are restricted to $\bq\to0$ and $\bq\simeq\bK$. To access the steady state at any wavevector, we solve the $4\times4$ Lyapunov equation (Appendix~\ref{Lyapunovequations}) exactly for each allowed $\bq$ of the lattice. Real-space correlations follow from the inverse Fourier transform. In particular, we consider the spatial velocity correlations, $\langle\mathbf v_\bn\cdot\mathbf v_{\mathbf 0}\rangle=\frac1N\sum_\bq\langle\tilde{\mathbf v}_\bq\cdot\tilde{\mathbf v}_{-\bq}\rangle e^{i\bq\cdot\bR^0_\bn}$, the kinetic temperature, $T_{\rm kin}=\frac{m}{2}\langle v^2\rangle$, and the angular momentum per particle, $L=m\langle u^x_\bn v^y_\bn-u^y_\bn v^x_\bn\rangle$. In equilibrium, $\langle\tilde{\mathbf v}_\bq\cdot\tilde{\mathbf v}_{-\bq}\rangle=2T/m$ for any $\bq$, so that velocities are uncorrelated in space, $T_{\rm kin}=T$ and $L=0$. Any deviation from these values is therefore a nonequilibrium effect.

Figure~\ref{fig:velocity} shows these quantities for $C_1/C_0=0.5$, for which the instability at $K$ occurs at $B/\gamma=1.5$. For a field opposite to the odd chirality ($B/\gamma=-1$, Fig.~\ref{fig:velocity}(a)), the velocity spectrum deviates only weakly from equipartition, with a broad maximum, $\langle\tilde{\mathbf v}_\bq\cdot\tilde{\mathbf v}_{-\bq}\rangle\approx1.2\times2T/m$, near the zone boundary. Increasing $B$, the spectrum develops peaks at the six corners of the Brillouin zone (Figs.~\ref{fig:velocity}(b) and (c)), which grow by more than one order of magnitude on approaching the instability, while remaining close to the equilibrium value at small $q$ (Fig.~\ref{fig:velocity}(d)). In real space (Fig.~\ref{fig:velocity}(e)), velocity correlations are short ranged for $B/\gamma=1$ but extend over many lattice spacings close to the instability. They form a $\sqrt3\times\sqrt3$ pattern, the real-space counterpart of the peaks at $K$, in which the velocity direction rotates by $2\pi/3$ between neighbouring sites, producing vortex-like structures around the triangular cell of the lattice. The corresponding correlation length, extracted from the Ornstein--Zernike form $\langle\tilde{\mathbf v}_\bq\cdot\tilde{\mathbf v}_{-\bq}\rangle\simeq A/[1+\xi^2|\bq-\bK|^2]$ (Fig.~\ref{fig:velocity}(f)), diverges as $\xi\sim(B_c-B)^{-1/2}$ at the instability, $B_c/\gamma=[1-\frac{w}{2}(C_1/C_0)^2]/(C_1/C_0)$. The magnetic field thus controls the range of the velocity correlations, which, in the absence of odd elasticity, would vanish identically.

Odd elasticity alone already drives the crystal out of equilibrium. The kinetic temperature exceeds the bath temperature, $T_{\rm kin}>T$, and the crystal acquires a finite angular momentum also for $B=0$ (Figs.~\ref{fig:velocity}(g) and (h)), consistently with the energy balance of Appendix~\ref{app_energetic_balance}. A magnetic field aligned with the odd chirality strongly enhances both effects: for $C_1/C_0=0.5$, $T_{\rm kin}/T$ increases from $1.18$ at $B=0$ to $1.46$ at $B/\gamma=1$. Approaching the instability, both $T_{\rm kin}$ and $|L|$ grow sharply and eventually diverge, logarithmically in two dimensions, because of the critical fluctuations at $\bK$.
Section~\ref{sec:structurefactor} translates the long-wavelength results of this Section into the static structure factor $S(\bq)$, the quantity accessed directly in scattering experiments.

\begin{figure*}[t]
\centering
\includegraphics[width=\textwidth]{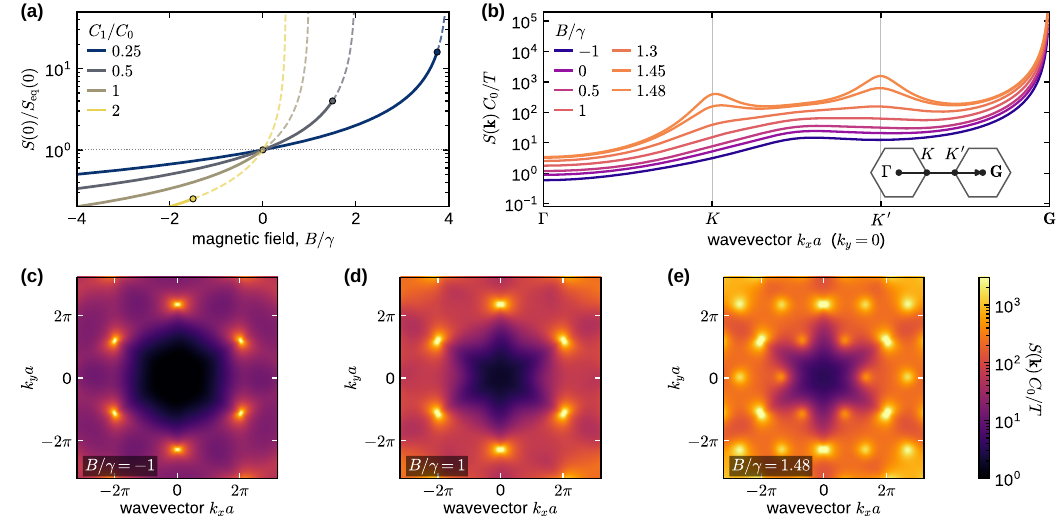}
\caption{\textbf{Static structure factor.}
(a) Forward-scattering intensity normalized by its equilibrium value, $S(0)/S_{\rm eq}(0)$, Eq.~\eqref{eq:S0},
as a function of $B/\gamma$ for several $C_1/C_0$. Solid lines correspond to a linearly stable crystal, while dashed lines show the extension of the curves beyond the onset of the instability at $\bK$, which is marked by dots.
(b) One-phonon structure factor $S(\bk)=k_\alpha k_\beta\langle \tilde{u}^\alpha_\bk \tilde{u}^\beta_{-\bk}\rangle$ along
the straight path $\Gamma\to K\to K'\to\mathbf G$ with $k_y=0$ (inset), where $K'=(8\pi/3a,0)$ is the corner of
the adjacent Brillouin zone and $\mathbf G=(4\pi/a,0)$ a reciprocal lattice vector, for several $B/\gamma$. The displacement covariance entering $S(\bk)$ is obtained from the exact solution of Eq.~\eqref{eq:lyap4x4} (see Appendix~\ref{Lyapunovequations}).
(c)--(e) $S(\bk)$ in reciprocal space for $B/\gamma=-1$ (c), $1$ (d) and $1.48$ (e), on a common
logarithmic color scale. The color bar shown next to (e) applies to panels (c), (d) and (e). Bragg delta peaks and the Debye--Waller
factor are omitted. The remaining parameters are: $w=9C_0/(m\gamma^2)=2$ in all panels, and $C_1/C_0=0.5$ in (b)--(e).}
\label{fig:structure}
\end{figure*}

\subsection{Static structure factor}
\label{sec:structurefactor}

The long-wavelength displacement fluctuations of Sec.~\ref{Staticproperties} translate directly into predictions for the static structure factor $S(\bq)$, which is typically measured in X-ray, neutron, or light scattering experiments~\cite{ashcroft1976solid,weling1983phonon}. Two distinct limits are physically relevant, and both are worked out in full in Appendix~\ref{staticSk}: the diffuse scattering surrounding a Bragg peak, $\bk=\mathbf G+\bq$ with $\mathbf G\neq0$ a reciprocal lattice vector and $|\bq|\ll|\mathbf G|$, and the true forward-scattering (compressibility) limit $\bk=\bq\to0$, with no reciprocal lattice vector involved. Although both arise from the same underlying displacement correlations, they are governed by different linearizations of the microscopic density and are not  obtained from one another by taking $\mathbf{G}\to0$
 (see the remark at the end of 
Appendix~\ref{staticSk}).
Near a Bragg peak, the diffuse intensity is
\begin{equation}
S(\mathbf G+\bq) = \frac{8G^2T}{3C_0}\,\frac{1}{q^2}\,\frac{1}{1-\frac{B}{\gamma}\frac{C_1}{C_0}}\left[1-\frac{2}{3}\cos^2\theta\right],
\label{eq:SGq_final}
\end{equation}
where, for concreteness, we have taken one of the smallest reciprocal lattice vectors, $\mathbf G=\frac{2\pi}{a}(1,1/\sqrt3)$, and $\theta$ is the angle between $\bq$ and $\mathbf G$, $\cos\theta=\hat{\mathbf G}\cdot\hat{\bq}$, so that $\mathbf G\cdot\hat{\bq}=G\cos\theta$. In the forward-scattering limit of vanishing wavevector, $\mathbf{q}=0$, we obtain
\begin{equation}
S(0) = \frac{8}{9}\,\frac{T}{C_0}\,\frac{1}{1-\frac{B}{\gamma}\frac{C_1}{C_0}}.
\label{eq:S0}
\end{equation}
Both expressions share the same $\bigl(1-\tfrac{B}{\gamma}\tfrac{C_1}{C_0}\bigr)^{-1}$ renormalization by the magnetic field and odd elasticity, the same factor that renormalizes the longitudinal and transverse displacement correlations for vanishing wavevector.
However, they probe different physics: Eq.~\eqref{eq:SGq_final} is an algebraic Bragg singularity, $S\sim1/q^2$, with amplitude set by $G^2$ and an angular dependence reflecting the anisotropy $\Sigma_{TT}=3\Sigma_{LL}$; Eq.~\eqref{eq:S0} is finite and independent of direction. In equilibrium, it would be fixed by the compressibility through the sum rule (Appendix~\ref{staticSk}); here, the factor $\bigl(1-\frac{B}{\gamma}\frac{C_1}{C_0}\bigr)^{-1}$ measures the violation of this sum rule, since the compressibility, determined by the static response, is not affected by the magnetic field. They also diverge for different reasons: Eq.~\eqref{eq:SGq_final} diverges as $\bq\to0$ at any fixed Bragg point, which is not an instability but the ordinary phonon divergence of a 2D crystal, a kinematic consequence of broken translational symmetry; Eq.~\eqref{eq:S0}, instead, diverges as $BC_1/(\gamma C_0)\to1$, signalling a long-wavelength dynamical instability, of different nature from the finite-wavevector instability of Sec.~\ref{sec:Kpoint_instability} (as discussed in Sec.~\ref{Staticproperties}, for $w>0$ this long-wavelength instability is reached only after the onset of the instability at $\bK$). The divergence of $S(0)$ is therefore a dynamical signature: it reflects the vanishing relaxation rate of the long-wavelength compressional modes, rather than a vanishing bulk modulus.

Equation~\eqref{eq:SGq_final} displays three main features: i) {\it Algebraic Bragg singularity}. As expected for a two-dimensional crystal, the diffuse intensity diverges as $S(\mathbf G+\mathbf q)\sim1/q^2$ on approaching the Bragg point.
ii) {\it Angular anisotropy.} The diffuse intensity depends on the angle between the scattering vector and the reciprocal lattice vector because longitudinal and transverse fluctuations enter with different weights, in the ratio $1:3$, reflecting the factor-of-three asymmetry $\Sigma_{TT}=3\Sigma_{LL}$ found earlier.
iii) {\it Magnetic/odd renormalization.} The Lorentz force and odd elasticity enter only through the overall amplitude, via the same factor $\bigl(1-\frac{B}{\gamma}\frac{C_1}{C_0}\bigr)^{-1}$ that enhances the long-wavelength displacement fluctuations (Sec.~\ref{Staticproperties}); the angular dependence is unchanged in this (isotropic continuum, leading-order) limit. As discussed in Sec.~\ref{Staticproperties}, this factor does not reflect a modification of the elastic moduli. Its origin is dynamical: during the relaxation of a compressional fluctuation, the odd elastic force generates a transverse particle motion, which the Lorentz force rotates into an additional force acting back on the displacement field. Depending on the sign of $BC_1$, this feedback slows down or speeds up the relaxation of the compressional mode, and thus enhances or suppresses its fluctuations. Neither the Lorentz force, which performs no work, nor the odd force alone, which does not affect the equal-time fluctuations at long wavelength, can produce this effect, which originates from the interplay between odd elasticity, Lorentz dynamics, and viscous dissipation.

\vskip8pt
\noindent
\textbf{Structure factor at arbitrary wavevector --} The two limits above are special cases of the one-phonon (harmonic) structure factor $S(\bk)=k_\alpha k_\beta\langle \tilde{u}^\alpha_\bk \tilde{u}^\beta_{-\bk}\rangle$, obtained from Eq.~\eqref{eq:Sk_master} by linearizing $e^{-i\bk\cdot\mathbf u_n}$ and discarding the Bragg peaks
(and the Debye-Waller factor). Here the displacement covariance is periodic in reciprocal space, while the prefactor $k_\alpha k_\beta$ is not, so $S(\bk)$ grows with $|\bk|$ from one Brillouin zone to the next. This expression reduces to Eq.~\eqref{eq:SGq_general} near a Bragg peak, $\bk=\mathbf G+\bq$ with $|\bq|\ll|\mathbf G|$, and to Eq.~\eqref{eq:Sq_def} for $\bk\to0$. Using the exact displacement covariance, it can be evaluated at any $\bk$ (Fig.~\ref{fig:structure}). Figure~\ref{fig:structure}(a) shows the forward-scattering intensity, Eq.~\eqref{eq:S0}, as a function of the magnetic field for several values of the odd elasticity. $S(0)$ increases (decreases) for $BC_1>0$ ($BC_1<0$), and the larger the odd elasticity, the stronger its dependence on $B$. Its divergence, however, always occurs beyond the onset of the instability at $\bK$ (dots). Figures~\ref{fig:structure}(b)--(e) show how this instability appears in a scattering experiment. For $B/\gamma=-1$, the diffuse intensity is concentrated around the Bragg peaks, with the anisotropic shape predicted by Eq.~\eqref{eq:SGq_final}. Increasing $B$, the overall intensity increases and new diffuse spots appear at the zone corners, $\bk=\mathbf G+\bK$, between the Bragg peaks. Close to the instability they reach an intensity comparable to the Bragg wings (Fig.~\ref{fig:structure}(b)). These superlattice spots are the scattering signature of the $\sqrt3\times\sqrt3$ fluctuations of Sec.~\ref{sec:velocitycorrelations}, and provide a direct experimental probe of the finite-wavevector instability.

\section{Time-Dependent Correlations}
\label{Timedependentcorrelations}
Equal-time correlations measure how large the fluctuations of the crystal are, but not how they evolve in time. This information is contained in two-time correlation functions, which are also the quantities most directly accessible in experiments, through dynamic light scattering or particle tracking. In this Section we show that the regimes identified in Sec.~\ref{Dynamicintheslattice} leave distinct fingerprints in the dynamics. The exceptional points separate a monotonic from an oscillatory relaxation of density fluctuations, and each of the two instabilities is announced by its own dynamical precursor: a slowing down of the long-wavelength relaxation for the instability at $\bq\to0$, and a sharp, weakly damped oscillation at finite frequency for the instability at $\bK$.

\subsection{Long-wavelength dynamics: monotonic versus oscillatory relaxation}
\label{sec:isf}
At small wavevectors, the dynamics is governed by the two soft modes $\lambda_\pm$ of Eq.~\eqref{diffusiveeigenvalues}, and the overdamped approximation of Sec.~\ref{Exceptionalline} applies. A natural observable is the intermediate scattering function, $F(\mathbf k,t)=\frac1N\langle\rho(\mathbf k,t)\rho(-\mathbf k,0)\rangle$, i.e.\ the time-dependent generalization of the static structure factor of Sec.~\ref{sec:structurefactor}. In the forward-scattering limit, $\mathbf k=\bq\to0$, only longitudinal displacements contribute, $F(\bq,t)=q^2C^{LL}_\bq(t)$, where $C^{LL}_\bq(t)$ is the two-time longitudinal displacement correlation derived in Appendix~\ref{app:twotime}; as in the static case, shear fluctuations are invisible in this limit. When the soft modes form a complex-conjugate pair, $r>p^2$, we obtain
\begin{equation}
F(\bq,t) = S(0)\; e^{-|t|/\tau(\bq)}\left[ \cos\!\bigl(\Omega(\bq)|t|\bigr) - \frac{p}{2\sqrt{r-p^2}}\,\sin\!\bigl(\Omega(\bq)|t|\bigr) \right],
\label{eq:Fqt_closed}
\end{equation}
where $S(0)$ is the forward-scattering intensity of Eq.~\eqref{eq:S0}, and
\begin{equation}
\tau^{-1}(q)=\nu p, \qquad \Omega(q)=\nu\sqrt{r-p^2},
\label{eq:tau_Omega}
\end{equation}
are the decay rate and the oscillation frequency of density fluctuations, with $\nu$, $p$ and $r$ given in Eq.~\eqref{eq:nupr}. Here, $\tau$ is the time over which the envelope of $F(\bq,t)$ decays, and $2\pi/\Omega$ is the period of its oscillations. They are directly related to the soft eigenvalues of Eq.~\eqref{diffusiveeigenvalues}, $\lambda_\pm=-\tau^{-1}\mp i\Omega$. Explicitly, the decay rate reads
\begin{equation}
\tau^{-1}=\frac{3}{4}\,\frac{C_0}{m\gamma}\,q^2\,\frac{1-\frac{B}{\gamma}\frac{C_1}{C_0}}{1+\frac{B^2}{\gamma^2}}\,,
\label{eq:tau_explicit}
\end{equation}
so that, in units of the diffusion rate $Dq^2$, with $D=C_0/(m\gamma)$, both $\tau^{-1}$ and $\Omega$ are independent of $q$. For real eigenvalues, $p^2>r$, the trigonometric functions are replaced by hyperbolic ones, Eq.~\eqref{eq:Fqt_overdamped}.

Equation~\eqref{eq:Fqt_closed} shows how the exceptional points of Sec.~\ref{Exceptionalline} become visible in the dynamics. On the diffusive side of the exceptional-point line, density fluctuations relax without oscillating, as in an ordinary overdamped crystal. Beyond it, they relax through damped oscillations, at a frequency set by the competition between odd elasticity and magnetic field. Figure~\ref{fig:dynamics}(d) illustrates this crossover for $C_1/C_0=0.5$, showing $F(\bq,t)$, normalized by its initial value, as a function of the rescaled time $q^2C_0t/(m\gamma)$. For a field opposite to the odd chirality, $B/\gamma=-1$, $F(\bq,t)$ decays monotonically to zero. Increasing $B$, the decay first develops a negative undershoot, which signals the onset of oscillations at the exceptional point, and then becomes slower and more oscillatory: for the largest fields, $F(\bq,t)$ reaches a deep negative minimum and relaxes over much longer times, as expected from the decrease of $\tau^{-1}$ discussed below. Since $F(\bq,t)$ can be measured by dynamic light scattering, the exceptional-point line can be located experimentally by tuning the magnetic field. Both $\tau^{-1}$ and $\Omega$ are proportional to $q^2$, so that $F(\bq,t)=S(0)\,f(q^2t)$ obeys the diffusive scaling of an ordinary overdamped solid, with dynamical exponent $z=2$: odd elasticity and magnetic field do not change the scaling, but only the shape of the scaling function $f$.

The decay rate also reveals how the long-wavelength instability is approached. According to Eq.~\eqref{eq:tau_explicit}, the magnetic field affects the relaxation in two different ways. The factor $1/(1+B^2/\gamma^2)$ is present also in passive crystals as follows from setting $C_1=0$ in Eq.~\eqref{eq:tau_explicit}): the Lorentz force deflects the particles and reduces their mobility along the restoring force, slowing down the relaxation without changing the equal-time fluctuations. The factor $1-BC_1/(\gamma C_0)$, instead, exists only in the presence of odd elasticity and is the same factor that controls the static fluctuations, Eq.~\eqref{eq:S0}. For $BC_1>0$, long-wavelength fluctuations thus become larger and slower at the same time, and their relaxation rate vanishes at $BC_1/(\gamma C_0)=1$. The long-wavelength instability is therefore announced by a dynamical slowing down, reminiscent of the critical slowing down near a continuous phase transition, although here it occurs out of equilibrium and leads to a dynamical instability. Since the oscillation frequency $\Omega$ remains finite at this point, close to the threshold $F(\bq,t)$ displays weakly damped oscillations at long times. For finite inertia, this point lies beyond the onset of the instability at $\bK$ (dashed lines in Fig.~\ref{fig:dynamics}(e)), consistently with Sec.~\ref{Staticproperties}. For $BC_1<0$, the opposite occurs: the field speeds up the relaxation of the long-wavelength modes and, at moderate strength, suppresses their oscillations.

These two effects are visible in Fig.~\ref{fig:dynamics}(e), which shows $\tau^{-1}$ and $\Omega$ in units of $Dq^2$ as a function of $B/\gamma$, for three values of $C_1/C_0$. The decay rate (top) is maximal at a moderately negative field, where the acceleration due to $BC_1<0$ is balanced by the reduced mobility, and decreases on both sides: slowly for $B<0$, where the reduced mobility dominates, and rapidly for $B>0$, where it vanishes linearly at $B/\gamma=C_0/C_1$. The oscillation frequency (bottom) vanishes in a window of fields where the dynamics is purely relaxational, and grows as a square root beyond the exceptional points that bound this window; as the odd elasticity increases, the window shifts towards negative fields.

\subsection{Finite-wavevector dynamics: propagating modes and persistent oscillations}
\begin{figure*}[t]
\centering
\includegraphics[width=\textwidth]{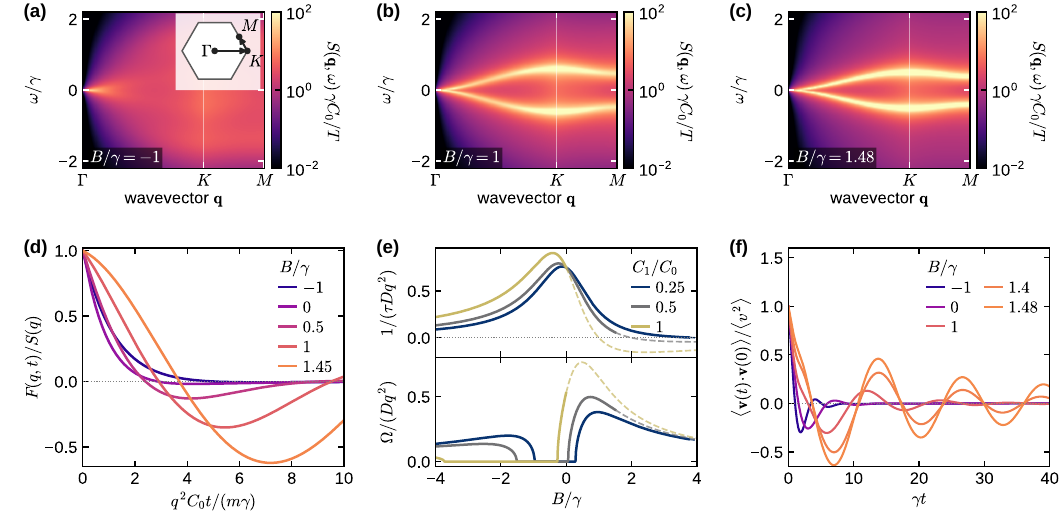}
\caption{\textbf{Dynamics.}
(a)--(c) Dynamic structure factor $S(\bq,\omega)=q_\alpha q_\beta C^{\alpha\beta}_\bq(\omega)$, Eq.~\eqref{eq:CABresult} (see Appendix~\ref{app:frequency}),
along the path $\Gamma\to K\to M$ (inset), for $B/\gamma=-1$ (a), $1$ (b) and $1.48$ (c), on a common
logarithmic color scale. (d) Intermediate scattering function $F(q,t)/S(q)$, Eq.~\eqref{eq:Fqt} (see Appendix~\ref{app:twotime}), with the time-dependent longitudinal correlation computed from Eq.~\eqref{eq:timecorr} (see Appendix~\ref{Lyapunovequations}),
as a function of the scaled time $q^2C_0t/(m\gamma)$, for several $B/\gamma$. (e) Decay rate $1/\tau$ (top) and
oscillation frequency $\Omega$ (bottom) of the long-wavelength modes, Eqs.~\eqref{eq:tau_Omega} and \eqref{eq:tau_explicit}, obtained in the overdamped approximation and shown in units of $Dq^2$ with $D=C_0/(m\gamma)$, as a function of $B/\gamma$ for several $C_1/C_0$. Dashed lines show the extension of the curves beyond the onset of the instability at $\bK$. 
(f) Single-particle velocity autocorrelation $\langle\mathbf v(t)\cdot\mathbf v(0)\rangle/\langle v^2\rangle$, Eq.~\eqref{eq:vacf} (see Appendix~\ref{Lyapunovequations}), for several $B/\gamma$.
All results, except panel (e), are obtained from the full underdamped linear dynamics. The remaining parameters are $w=9C_0/(m\gamma^2)=2$ in all panels; $C_1/C_0=0.5$ in (a)--(d) and (f), for which the instability at $K$ occurs at $B/\gamma=1.5$; $qa=0.3$, with $\bq$ along $x$, in (d). The sum over wavevectors in (f) includes all the wavevectors of a periodic crystal of $64\times64$ particles.}
\label{fig:dynamics}
\end{figure*}

At finite wavevectors inertia cannot be neglected, and we compute the dynamic structure factor, $S(\bq,\omega)=q_\alpha q_\beta C^{\alpha\beta}_\bq(\omega)$, from the full underdamped linear dynamics, Eq.~\eqref{eq:CABresult} (Appendix~\ref{app:frequency}). This quantity resolves the frequency content of density fluctuations at each wavevector and is measured in inelastic scattering experiments. Figures~\ref{fig:dynamics}(a)--(c) show $S(\bq,\omega)$ for $C_1/C_0=0.5$ along the path $\Gamma\to K\to M$ of the Brillouin zone, as a function of the frequency $\omega/\gamma$, on a logarithmic color scale; bright regions indicate the frequencies at which density fluctuations of a given wavevector are mostly excited. For a field opposite to the odd chirality (Fig.~\ref{fig:dynamics}(a)), the spectrum is dominated by a broad peak at $\omega=0$: fluctuations relax without a characteristic frequency. For a field aligned with the odd chirality (Fig.~\ref{fig:dynamics}(b)), the spectral weight moves to two branches at $\pm\omega(\bq)$, which reach their maximal frequency close to $\bK$: density fluctuations now propagate as damped waves. On approaching the instability at $\bK$ (Fig.~\ref{fig:dynamics}(c)), these branches become sharp, because the damping of the corresponding modes vanishes. The undamped mode predicted in Sec.~\ref{sec:Kpoint_instability} thus appears as a narrow resonance at finite frequency and finite wavevector, whose growth and narrowing are a direct precursor of the instability.

The same mode also governs the motion of single particles. The velocity autocorrelation, $\langle\mathbf v(t)\cdot\mathbf v(0)\rangle$, normalized by $\langle v^2\rangle$ and shown in Fig.~\ref{fig:dynamics}(f) as a function of $\gamma t$, decays within a few inertial times $\gamma^{-1}$ for a field opposite to the odd chirality, but develops long-lived oscillations close to the instability, at the frequency of the weakly damped mode at $\bK$. Each particle thus keeps memory of the collective rotation of its neighbours: these oscillations are the temporal counterpart of the vortex pattern of Fig.~\ref{fig:velocity}(e), in which velocities rotate coherently around the triangular plaquettes.

In summary, the two instabilities identified in Secs.~\ref{Dynamicintheslattice} and~\ref{Staticproperties} are announced by different dynamical precursors: a slow long-wavelength relaxation, whose rate vanishes as $BC_1/(\gamma C_0)\to1$, and a sharp finite-frequency resonance at $\bK$, accompanied by persistent oscillations of single-particle velocities. Both are controlled by the magnetic field and are accessible to scattering and particle-tracking experiments.

\section{Conclusions}
\label{Conclusions}

The Lorentz force is non-dissipative: it performs no work and, by itself, does not modify the stationary fluctuations of an equilibrium system. Nevertheless, when combined with odd elasticity in odd active crystals, it profoundly changes both the dynamics and the static correlations, because it couples to the circulating currents generated by the odd forces. 
In particular, the magnetic force generates a new instability at vanishing wavevector, which is absent in equilibrium solids as well as in odd active crystals without a magnetic force. Indeed, this is different from the finite-wavevector instability induced by odd elasticity alone \cite{caprini2025odd} which requires no magnetic field.

Moreover, the Lorentz force modifies the spectrum of the non-Hermitian dynamical matrix generated by odd elasticity, favoring oscillations by modulating the exceptional points and promoting modes with vanishing effective damping near the edge of the Brillouin zone. These modes occur at a finite wavevector and correspond to a propagating, non-dissipative branch, which suggests a stochastic analogue of a finite-$q$ oscillatory (Hopf-like) instability~\cite{walgraef2012spatio} in which thermal noise excites a persistent oscillatory mode with unbounded amplitude in the absence of nonlinear saturation. This mechanism produces large finite-$q$ fluctuations and eventually instabilities that can be tuned through a gyroscopic force. 
The exact solution of the linear dynamics at arbitrary wavevector shows how the finite-wavevector instability manifests in observable quantities: approaching the instability, velocities become correlated over a diverging length and form a pattern of vortex-like structures. Odd elasticity alone raises the kinetic temperature above the bath temperature and generates a finite angular momentum, and a magnetic field aligned with the odd chirality strongly amplifies both effects. The same fluctuations produce diffuse superlattice spots at the Brillouin-zone corners in the static structure factor, two propagating branches that sharpen as their damping vanishes in the dynamic structure factor, and long-lived oscillations of the velocity autocorrelation. These signatures, accessible through particle tracking and scattering experiments, provide direct probes of the magnetic control of odd elasticity.

Our results for the longitudinal and transverse displacement correlations have consequences for the interpretation of fluctuation-based measurements of the elastic moduli of an odd active crystal in the presence of inertia and magnetic field. Indeed, from the expressions \eqref{eq:long_trans_correlations}, one can define, through the equilibrium relation between fluctuations and moduli, the effective shear and effective bulk elastic moduli, $\mu_{\mathrm{eff}}$ and $B_0^{\mathrm{eff}}$, given by
\begin{subequations}
\begin{flalign}
\mu_{\mathrm{eff}} &= \lim_{q \to 0} \frac{2T}{\sqrt3a^2 q^2 \Sigma_{TT}(q)} = \frac{\sqrt3}{4} C_0 \left(1 - \frac{B}{\gamma}\frac{C_1}{C_0}\right), \\
B_0^{\mathrm{eff}} + \mu_{\mathrm{eff}} &= \lim_{q \to 0} \frac{2T}{\sqrt3a^2 q^2 \Sigma_{LL}(q)} = 3 \mu_{\mathrm{eff}} \,,
\end{flalign}
\end{subequations}
where $\sqrt3a^2/2$ is the area per particle, which converts the fluctuations per particle into moduli per unit area. These fluctuation-based moduli differ from the mechanical ones, Eq.~\eqref{eq:moduli_mapping}, which are not affected by the magnetic field: their reduction for $BC_1>0$ signals a nonequilibrium enhancement of the fluctuations, rather than a softening of the crystal. A comparison between static response and displacement fluctuations would therefore directly reveal the magnetic control of odd elasticity.
In the overdamped limit, at vanishing magnetic field ($B=0$), the antisymmetric part of the dynamical matrix of displacement interactions does not contribute to dissipation, while in the absence of odd elasticity the magnetic field does not contribute to the fluctuations either; 
when both are present, their combined action generates the renormalization factor $\bigl(1-\frac{B}{\gamma}\frac{C_1}{C_0}\bigr)^{-1}$ of Eqs.~\eqref{eq:long_trans_correlations} and~\eqref{eq:Teff}.
When both mechanisms are simultaneously present, their combined action generates a term that does contribute to the fluctuations. An analogous scenario occurs in odd fluids, where odd viscosity \cite{avron1998odd, fruchart2023odd, reichhardt2022active, mecke2023simultaneous, hosaka2023lorentz, markovich2024nonreciprocity} and odd torque combine to generate an instability of the uniform fluid~\cite{marini2026emergent}.

Consequently, our findings suggest magnetic fields as a powerful external control parameter for tuning the dynamics and stability of odd elastic materials, and pave the way towards the experimental investigations of non-Hermitian mechanics in active metamaterials \cite{chen2021realization}, chiral active vibrobots \cite{scholz2021surfactants, siebers2023exploiting, tiwari2026reentrant, caprini2026active}, or chiral colloids \cite{yan2014rotating, kokot2017active, massana2021arrested,bililign2022motile}. 
The Lorentz force does not generate nonequilibrium behavior by itself; rather, it acts as a probe and controller of the circulating currents generated by odd elasticity.
While the effect of a gyroscopic force on single chiral self-propelled particles has already been addressed, and that on odd crystals has been discussed here, one can wonder whether a Lorentz force could reshape collective phenomena of chiral active particles, and, specifically, edge currents \cite{van2016spatiotemporal,adorjani2024phase,langford2025phase,kuroda2026designing,metzger2026equationstateedgeflow,wang2026edge,caprini2026active} and the bubble phase \cite{caprini2025bubble} (BIO) recently observed in chiral liquids with odd interactions. The instability observed at small wavevector in the crystal case suggests that a gyroscopic force could transform bubbles with finite size into a macroscopic phase separation in dense fluid systems. 
In principle, the Lorentz force can influence the collective phenomena observed in chiral self-propelled fluids \cite{levis2019activity, huang2020dynamical, reichhardt2019reversibility, zhang2020reconfigurable, mecke2024emergent, valecha2025active,caprini2025spontaneous, jeong2025crystallization, kalz2026reversal, shen2026emergent}, including self-reverting vorticity in dense clusters \cite{caprini2024self}, as well as hyperuniformity \cite{zhang2022hyperuniform,kuroda2023microscopic,maire2025hyperuniformity,liu2025hyperuniform} in the limit of vanishing persistence time. Similarly, one can wonder whether a gyroscopic force could modify the chimera states~\cite{kreienkamp2025synchronization} and the rotating flocking  induced by the competition between chirality and alignment interactions~\cite{musacchio2026circling}.
 
Finally, we comment on the experimental realization of the model. For colloidal or granular particles, the Lorentz force produced by realistic magnetic fields is negligible compared with friction, $B/\gamma\ll1$. However, Eq.~\eqref{ALangevin1} only requires a force proportional and perpendicular to the particle velocity, a structure shared by other mechanisms. The most direct one is the Coriolis force: in a frame rotating with angular velocity $\Omega_{\rm rot}$ around $\hat{\mathbf z}$, each particle experiences the force $-2m\,\Omega_{\rm rot}\hat{\mathbf z}\times\mathbf v$, identical to the Lorentz force with $B=-2\Omega_{\rm rot}$, while the centrifugal force has to be compensated, for instance by a suitably curved substrate, or is negligible for small systems. Chiral active granular particles, such as vibrobots~\cite{scholz2018rotating, carrillo2025depinning, caprini2026active} or air-driven spinners \cite{farhadi2018dynamics, lopez2022chirality, vega2022diffusive}, on a rotating platform would thus realize the present model, with $B/\gamma$ tuned by the rotation rate. Gyroscopic forces of the same form also arise in mechanical metamaterials made of spinning elements, which could be combined with the non-reciprocal couplings of active metamaterials~\cite{brandenbourger2019non, chen2021realization,veenstra2024non}. A genuine Lorentz force can instead act on charged grains in complex (dusty) plasmas \cite{kahlert2012magnetizing}, which form weakly damped two-dimensional crystals with non-reciprocal interactions, although very strong fields are needed to magnetize the heavy grains, so that the rotating-frame strategy is again the most promising. In these systems, the quantities discussed here -- velocity correlations, kinetic temperature, angular momentum and structure factors -- are directly accessible by particle tracking and the magnetic field can be tuned to explore both instabilities.
 
 \section*{Data availability statement}
\noindent
All data that support the findings of this study are included within the article.

\appendix

\section{Direct and Reciprocal Lattice}
\label{app:lattice}
In this Appendix, we summarize direct and reciprocal vectors on the lattice setting the notation for the rest of the calculations.
In our chiral active crystal, the particles at rest sit on the nodes of a triangular $L\times L$ lattice, with $N=L^2$ sites:
\begin{equation}
\bR^0_\bn = n_1 \mathbf{a}_1 + n_2 \mathbf{a}_2,
\end{equation}
with primitive vectors
%\begin{subequations}
\begin{align}
\mathbf{a}_1 &= \frac{a}{2} \hat{\mathbf{x}} - \frac{a\sqrt{3}}{2} \hat{\mathbf{y}},&
\mathbf{a}_2 &= \frac{a}{2} \hat{\mathbf{x}} + \frac{a\sqrt{3}}{2} \hat{\mathbf{y}}.
\end{align}
%\end{subequations}
The displacement can be defined as $\rbu_\bn \equiv \bR_\bn - \bR^0_\bn$ and the system allows the following wavevectors:
\begin{equation}
\bq= {\bq}(l_1,l_2)= \frac{l_1}{L} {\bf G}_1+\frac{l_2}{L} {\bf G}_2, \qquad 0<l_1,l_2\leq L,
\end{equation}
with
\begin{align}
{\bf G}_1 &=\frac{ 2 \pi}{a} \Bigl( \hat{\bf x}- \frac{1}{\sqrt 3} \hat{\bf y}\Bigr), &
{\bf G}_2 &= \frac{ 2 \pi}{a} \Bigl( \hat{\bf x}+\frac{1}{\sqrt 3} \hat{\bf y}\Bigr).
\end{align}
With these definitions, we can identify the unit vectors to the six nearest neighbors as
\begin{align}
{\bf e}_{1,4}&=\pm(1,0)=\pm({\bf a}_1 +{\bf a}_2), &
{\bf e}_{2,5}&=\pm\Bigl(\frac{1}{2},\frac{ \sqrt 3}{2}\Bigr) =\pm{\bf a}_2, &
{\bf e}_{3,6}&=\pm\Bigl(-\frac{1}{2},\frac{ \sqrt 3}{2}\Bigr)=\mp{\bf a}_1.
\end{align}
and the six corners of the first Brillouin zone as
\begin{align}
{\bf K}_{1,4}&=\pm\frac{2\pi}{3a}(2,0), &
{\bf K}_{2,5}&=\pm\frac{2\pi}{3a}(1,\sqrt{3}), &
{\bf K}_{3,6}&=\pm\frac{2\pi}{3a}(-1,\sqrt{3}).
\end{align}

\section{Pairwise Forces}
\label{Linearization}

The pairwise forces are of two types: central forces, derivable from a potential $U$, and transverse forces, which are not. We take
\begin{equation}
U(|\bR_\bn-\bR_\mb|)=\frac{C_0}{2}(|\bR_\bn-\bR_\mb|-\ell)^2.
\end{equation}
Differentiating and linearizing with respect to the displacements, with ${\bf e}_{\bn\mb}\equiv(\bR_\bn^0-\bR_\mb^0)/|\bR_\bn^0-\bR_\mb^0|$, $\delta\bu_{\bn\mb}\equiv\bu_\bn-\bu_\mb$, $\ell=|\bR_\bn^0-\bR_\mb^0|$:
\begin{equation}
{\bf F}^{even}_{\bn\mb}\approx -C_0 ({\bf e}_{\bn\mb} \cdot \delta \bu_{\bn\mb}) {\bf e}_{\bn\mb} -\frac{C_0}{\ell} \Bigl[({\bf e}_{\bn\mb} \cdot \delta \bu_{\bn\mb}) {\delta \bu}_{\bn\mb} +\frac{1}{2} \Bigl( 2 ({\bf e}_{\bn\mb} \cdot \delta \bu_{\bn\mb})^2 - (\delta \bu_{\bn\mb})^2\Bigr) {\bf e}_{\bn\mb} \Bigr].
\label{forzadiVitelli2}
\end{equation}
The first term is the harmonic projection along the bond; the second term contains the leading geometric nonlinearities. The odd contribution is
\begin{equation}
{\bf F}^{\rm odd}_{\bn \mb} \approx -C_1\, ({\bf e}_{\bn\mb} \cdot \delta \bu_{\bn\mb}) \, \hat{{\bf z}}\times {\bf e}_{\bn\mb}.
\end{equation}
Since $\delta\bu_{\bn\mb}=-(\rbu_\mb-\rbu_\bn)$, the harmonic term of Eq.~\eqref{forzadiVitelli2} and the odd force above coincide with Eqs.~\eqref{forcevitelli1} and \eqref{forcevitelli2} of the main text, respectively.
The harmonic approximation requires the root-mean-square relative displacement between nearest neighbours, $\sqrt{\langle|\rbu_\bn-\rbu_\mb|^2\rangle}$, to remain much smaller than the lattice spacing $\ell$. The nonlinear terms in Eq.~\eqref{forzadiVitelli2} are neglected in the rest of the paper.

%%%%%%%%%%%%%%%%%%%%%%%%%%%%%%%%%%%%%%%%%%%%%%
\section{Fourier transform and energy balance}
\label{app_energetic_balance}

We introduce the discrete spatial Fourier transforms and inverse Fourier transform of the velocity $\mathbf{v}_{\mathbf{n}}$ or any dynamical variables -- including displacement vector $\mathbf{u}_{\mathbf{n}}$ and noise $\boldsymbol{\xi}_{\mathbf{n}}$ -- as
\begin{subequations}
\begin{flalign}
&\vvt_{\bq}(t)=
\frac{1}{\sqrt{N}}
\sum_{\bn}
\mathbf{v}_{\bn}(t)\,
e^{-i\bq\cdot\bR^0_{\bn}}, \label{eq:FT_def}\\
&\vv_{\bn}(t)=
\frac{1}{\sqrt{N}}
\sum_{\bq}
\vvt_{\bq}(t)\,
e^{i\bq\cdot\bR^0_{\bn}} \,,
\end{flalign}
\end{subequations}
where $\mathbf{q}$ is two-dimensional wavevector and the transformed variable is denoted by a tilde.
We remark that for real-space displacements and velocities, $\tilde{\mathbf u}_{-\mathbf q}=\tilde{\mathbf u}_{\mathbf q}^{*}$ and $\tilde{\mathbf v}_{-\mathbf q}=\tilde{\mathbf v}_{\mathbf q}^{*}$.

By multiplying Eq.~\eqref{ALangevin1} by $\vvt_{-\bq}(t)$, symmetrizing the resulting expression, and averaging over the noise~\cite{caprini2025odd}, we obtain
\begin{equation}
\frac{d}{dt}
\Bigl\langle
\frac{m}{2}\,\vvt_\bq\cdot\vvt_{-\bq}
+
U^{\rm even}_\bq
\Bigr\rangle
+
m\gamma
\bigl\langle
\vvt_\bq\cdot\vvt_{-\bq}
\bigr\rangle
=
\bigl\langle
\tbfF^{\rm odd}_\bq\cdot\vvt_{-\bq}
\bigr\rangle
+
2\gamma T ,
\label{balanceequation}
\end{equation}
where $U^{\rm even}_\bq$ denotes the potential-energy contribution associated with mode $\bq$, such that  
$\frac{d}{dt} \langle U^{\rm even}_\bq \rangle= \bigl\langle
\tbfF^{\rm even}_\bq\cdot\vvt_{-\bq}
\bigr\rangle$.
The magnetic force gives a vanishing contribution to the balance equation, Eq.~\eqref{balanceequation}, since $\bigl\langle (\mathbf{B}\times \vvt_{\bq})\cdot\vvt_{-\bq}\bigr\rangle=0$ after symmetrization.
 Consequently, in the steady state, we have
\begin{equation}
\frac{m}{2T}
\bigl\langle
\vvt_\bq(t)\cdot\vvt_{-\bq}(t)
\bigr\rangle
-1
=
\frac{1}{2\gamma T}
\bigl\langle
\tbfF^{\rm odd}_\bq(t)\cdot\vvt_{-\bq}(t)
\bigr\rangle .
\label{energybalance}
\end{equation}
This balance equation implies that the magnetic field does not contribute to the energy budget, consistently with the fact that the Lorentz force performs no work, while odd forces produce deviations of the kinetic temperature from the bath temperature $T$. This is the result anticipated in Sec.~\ref{Oddelasticmodel}, and it explains why $T_{\rm kin}\neq T$ in Fig.~\ref{fig:velocity}(g) also at $B=0$.

\section{From microscopic force constants to continuum elastic moduli}\label{elasticmoduli}

At wavelengths much larger than the lattice spacing, the displacement field can be regarded as a smooth continuum field
\cite{landau2012theory,lautrup2011physics}. The corresponding equations of linear elastodynamics are
\begin{equation}
\rho\,\ddot u^\alpha
+\gamma \rho\,\dot u^\alpha-B\rho \epsilon^{\alpha\beta} \dot u^\beta
=
\nabla^\beta \sigma^{\alpha\beta}
+f^\alpha,
\label{eq:navierequation}
\end{equation}
where $\rho$ is the mass density, $f^\alpha$ is an external body force per unit volume, and $\sigma^{\alpha\beta}$ is the stress tensor.
Notice that the Lorentz term
 acts on each particle's  own velocity, with no dependence whatsoever on neighbors and therefore does not contribute to the stress tensor.
In two dimensions, the even (Cauchy) contribution to the stress tensor is
\begin{subequations}
\begin{align}
\sigma_e^{xx} &= (B_0+\mu)\,\nabla^x u^x +(B_0-\mu)\,\nabla^y u^y, \\
\sigma_e^{xy} = \sigma_e^{yx} &= \mu\,(\nabla^y u^x+\nabla^x u^y), \\
\sigma_e^{yy} &= (B_0+\mu)\,\nabla^y u^y +(B_0-\mu)\,\nabla^x u^x ,
\end{align}
\end{subequations}
while odd elasticity introduces the additional traceless, generally non-symmetric contribution
\begin{subequations}
\begin{align}
\sigma_o^{xx} &= -K_0(\nabla^x u^y+\nabla^y u^x), \\
\sigma_o^{xy} &= -A(\nabla^x u^x+\nabla^y u^y) +K_0(\nabla^x u^x-\nabla^y u^y), \\
\sigma_o^{yx} &= A(\nabla^x u^x+\nabla^y u^y) +K_0(\nabla^x u^x-\nabla^y u^y), \\
\sigma_o^{yy} &= K_0(\nabla^y u^x+\nabla^x u^y),
\label{continuuelasticity}
\end{align}
\end{subequations}
where $B_0,\mu$ are the conventional bulk and shear moduli, and $A,K_0$ are the odd bulk and odd shear moduli. Substituting $\sigma^{\alpha\beta}=\sigma_e^{\alpha\beta}+\sigma_o^{\alpha\beta}$ into Eq.~\eqref{eq:navierequation} and Fourier transforming,
\begin{equation}
-\omega^2 \rho\,u^\alpha -i\omega\gamma\rho\,u^\alpha+i\omega B\rho \epsilon^{\alpha\beta} u^\beta
=
-B_0 q^\alpha q^\beta u^\beta -\mu q^2 u^\alpha
+K_0 \epsilon^{\alpha\gamma} q^2 u^\gamma
+A \epsilon^{\alpha\gamma} q^\gamma q^\beta u^\beta
+f^\alpha.
\label{fouriernavier}
\end{equation}
Matching this to the long-wavelength expansion of the lattice dynamical matrix (Eqs.~\eqref{M52}-\eqref{M55}), with mass density $\rho=2m/(\sqrt3a^2)$, i.e.\ the particle mass divided by the area per particle of the triangular lattice, gives the correspondence between microscopic force constants and continuum elastic moduli,
\begin{equation}
\mu = \frac{B_0}{2} = \frac{\sqrt3}{4} C_0,
\qquad
K_0 = \frac{A}{2} = \frac{\sqrt3}{4} C_1 .
\label{eq:moduli_mapping}
\end{equation}
These are the conventional and odd moduli introduced in Sec.~\ref{Oddelasticmodel}. These relations are specific to the present triangular lattice with nearest-neighbour harmonic interactions.

The antisymmetric part of the stress tensor gives rise to a torque density,
\begin{equation}
\tau = \sigma_o^{xy} - \sigma_o^{yx} = -2A(\nabla^x u^x+\nabla^y u^y) = 2A\,\nabla\cdot\mathbf u.
\label{eq:torquedensity}
\end{equation}
Local compression or dilation generates an internal torque density: in ordinary elasticity ($A=0$), isotropic compression produces only a pressure variation, whereas in an odd elastic medium it simultaneously induces a local torque. This mechanism differs fundamentally from Cosserat or micropolar elasticity, where torque densities originate from independent rotational degrees of freedom: here the torque density is generated solely by the transverse non-conservative forces encoded in $C_1$, with no rotational degrees of freedom at all. While the continuum constitutive law depends only on the four elastic moduli above in the long-wavelength limit, the complete lattice dynamical matrix $M_\bq$ contains additional information at finite wavevector; in particular, its non-normal structure cannot be inferred from the continuum moduli alone.

%%%%%%%%%%%%%%%%%%%%%%%%%%%%%%%%%%%%%%%%%%%%%%%%
%%%%%%%%%%%%%%%%%%%%%%%%%%%%%%%%%%%%%%%%%%%%%%%%

\section{The Dynamical Matrix}
\label{matrixelements}

In Fourier representation, the dynamics read
\begin{equation}
\frac{d^2 \tilde{\mathbf u}_\bq}{dt^2}+ \gamma \frac{d \tilde{\mathbf u}_\bq}{dt}=
\frac{1}{m}
\Bigl({\bf F}^{\mathrm{even}}_{\bq} + {\bf F}^{\mathrm{odd}}_{\bq}\Bigr)
+ \,{\bf B}\times \frac{d \tilde{\mathbf u}_\bq}{dt}
+ \sqrt{\frac{2\gamma T}{m}}\,\boldsymbol{\xi}_\bq(t),
\label{ALangevin1fourier}
\end{equation}
where the $\alpha$ component of the total force $F^\alpha_\bq/m=M^{\alpha\beta}_\bq \tilde{u}^\beta_\bq$, can be expressed through the matrix $M^{\alpha\beta}_\bq$ with elements:
\begin{subequations}
\begin{align}
M^{xx}_\bq&=\frac{C_0}{2m}\Bigl(4 {\cal{C}}_{12}+{\cal{C}}_1+{\cal{C}}_2-6\Bigr) -\frac{C_1}{m} \frac{\sqrt 3}{2} \Bigl( {\cal{C}}_2-{\cal{C}}_1 \Bigr), \label{C3}\\
M^{yy}_\bq&= \frac{C_0}{m} \frac{ 3}{2} \Bigl( {\cal{C}}_2+{\cal{C}}_1 -2 \Bigr) +\frac{C_1}{m} \frac{\sqrt 3}{2} \Bigl( {\cal{C}}_2-{\cal{C}}_1\Bigr), \label{C4}\\
M^{xy}_\bq&= \frac{C_0}{m} \frac{\sqrt 3}{2} \Bigl( {\cal{C}}_2-{\cal{C}}_1 \Bigr) -\frac{C_1}{m} \frac{ 3}{2} \Bigl({\cal{C}}_2+{\cal{C}}_1 -2 \Bigr), \label{C5}\\
M^{yx}_\bq&= \frac{C_0}{m} \frac{\sqrt 3}{2} \Bigl( {\cal{C}}_2-{\cal{C}}_1 \Bigr) +\frac{C_1}{2m} \Bigl( 4 {\cal{C}}_{12}+{\cal{C}}_1+{\cal{C}}_2-6 \Bigr) \label{C6} \,.
\end{align}
\end{subequations}
Here, the constants ${\cal{C}}_1$, ${\cal{C}}_2$, and ${\cal{C}}_{12}$ have the following expressions:
\begin{subequations}
\begin{align}
{\cal{C}}_1&\equiv \cos({\bf q}\cdot {\bf a}_1)= \cos(q_x/2-\sqrt{3}/2\,q_y), \label{C7}\\
{\cal{C}}_2&\equiv\cos({\bf q}\cdot {\bf a}_2)=\cos(q_x/2+\sqrt{3}/2\,q_y), \label{C8}\\
{\cal{C}}_{12}&\equiv \cos({\bf q}\cdot ({\bf a}_1+ {\bf a}_2)) =\cos(q_x). \label{C9}
\end{align}
\end{subequations}
Equations~\eqref{C3}--\eqref{C9} define the matrix $\mathbf M_\bq$ of Eq.~\eqref{eq:fourier_space_eq}, and have been used for all the results at finite wavevector shown in Figs.~\ref{fig:spectrum}, \ref{fig:phasediagram}(b), \ref{fig:velocity}, \ref{fig:structure}(b)--(e) and \ref{fig:dynamics}(a)--(d),(f).
For a generic wavevector, $M_\bq^{xx}\neq M_\bq^{yy}$ and $M_\bq^{xy}\neq-M_\bq^{yx}$; the matrix takes an isotropic form only at the corners of the Brillouin zone.
In the long-wavelength limit and for $\bq=q\hat{\mathbf x}$, we obtain
\begin{subequations}
\begin{align}
M_\bq^{xx}&=-\frac{9 C_0}{8m} q^2, \label{M52}\\
M_\bq^{yy}&=-\frac{3 C_0}{8m} q^2, \label{M53}\\
M_\bq^{xy}&=\frac{3 C_1}{8m} q^2, \label{M54}\\
M_\bq^{yx}&=-\frac{9 C_1}{8m} q^2. \label{M55}
\end{align}
\end{subequations}
These expressions coincide with the long-wavelength matrix elements given in Sec.~\ref{Oddelasticmodel}. At the corners of the Brillouin zone ($\bq=\bK$, e.g. $\bK=(4\pi/3a,0)$), symmetry gives
\begin{subequations}
\begin{align}
\omega_\bK^2 &=-M^{xx}_\bK=-M^{yy}_\bK= \frac{9}{2}\frac{C_0}{m}, \label{mk2app}\\
\alpha_\bK^2 &=M^{xy}_\bK=-M^{yx}_\bK= \frac{9}{2}\frac{C_1}{m}. \label{mk3app}
\end{align}
\end{subequations}
which coincide with Eqs.~\eqref{mk2} and \eqref{mk3}.

\section{Derivation of the Dispersion Relation and its Long-Wavelength Expansions}
\label{app:spectrum_derivation}

This Appendix collects the algebraic steps behind the results quoted in Sec.~\ref{Dynamicintheslattice}.

\subsection{Fourier representation and the characteristic quartic equation}

We introduce a Fourier representation for the displacement field via the plane-wave ansatz
\begin{equation}
\bu_\bn(t)=\sum_\bq \tilde{\mathbf u}_\bq\, e^{i \bq \cdot \bR^0_\bn+\lambda t}.
\end{equation}
Substituting this ansatz into the linearized equation of motion~\eqref{ALangevin1} gives
\begin{equation}
\Bigl[
(\lambda^2 + \gamma\lambda) \mathbf{I} - \lambda \mathcal{B} - M_\bq
\Bigr]\tilde{\mathbf u}_\bq=0,
\label{eq:eigenvalueproblem}
\end{equation}
where $\mathcal{B}=B\epsilon$ is the antisymmetric matrix associated with the gyroscopic term, since $\epsilon$ is the two-dimensional Levi-Civita tensor. Non-trivial solutions require the vanishing of the characteristic determinant, yielding the quartic equation
\begin{equation}
P(\lambda)=\lambda^4 + 2\gamma \lambda^3
+ \left(\gamma^2 + B^2 - \Tr M_\bq\right) \lambda^2
- \left(\gamma\,\Tr M_\bq + B(M_\bq^{xy}-M_\bq^{yx})\right) \lambda
+ \det M_\bq=0,
\label{eqquartogrado_final}
\end{equation}
quoted in the main text. This equation governs the spectrum at arbitrary $\bq$, and reduces to the results discussed below in its long-wavelength limit and, in Appendix~\ref{Kpoint}, at the corners of the Brillouin zone. Equation~\eqref{eqquartogrado_final} has been used to compute the spectra of Fig.~\ref{fig:spectrum} and the acoustic frequency of Fig.~\ref{fig:phasediagram}(b).

\subsection{Long-wavelength expansion of $M_\bq$}

Expanding $M_\bq$ (Appendix~\ref{matrixelements}) to leading order in $q$ gives $\Tr M_\bq=-2\omega_\bq^2$ and $M_\bq^{xy}-M_\bq^{yx}=2\alpha_\bq^2$, with
\begin{equation}
\omega_\bq^2 =-\tfrac{1}{2}\Tr M_\bq \simeq \frac{3}{4}\frac{C_0}{m}q^2, \qquad
\alpha_\bq^2 = \tfrac{1}{2}(M_\bq^{xy}-M_\bq^{yx}) \simeq \frac{3}{4}\frac{C_1}{m}q^2, \qquad
\det(M_\bq)\simeq\frac{3}{4} \left[ (\omega_\bq^2)^2+ (\alpha_\bq^2)^2\right].
\label{eq:alphaq2def}
\end{equation}
At $q\to0$ exactly, Eq.~\eqref{eqquartogrado_final} reduces to $\lambda^2(\lambda^2+2\gamma\lambda+\gamma^2+B^2)=0$, giving two vanishing roots (double) and two complex roots $\lambda^{(3,4)}=-\gamma\pm iB$.

\subsection{Adiabatic elimination of the fast modes: derivation of Eq.~\eqref{diffusiveeigenvalues}}

At small but finite $q$, the two soft roots scale as $\lambda=O(q^2)$, while $\Tr M_\bq=O(q^2)$ and $\det M_\bq=O(q^4)$. Retaining only the leading order in $q$ for a root of this size, the quartic~\eqref{eqquartogrado_final} reduces to the quadratic
\begin{equation}
(\gamma^2+B^2)\lambda^2 + \bigl[2\gamma\omega_\bq^2-2B\alpha_\bq^2\bigr]\lambda + \det M_\bq = 0.
\end{equation}
Using Eq.~\eqref{eq:alphaq2def}, so that $\frac34 q^2=\nu(\gamma^2+B^2)$ with $\nu=\frac34 q^2/(\gamma^2+B^2)$, one finds $\gamma\omega_\bq^2-B\alpha_\bq^2=\nu(\gamma^2+B^2)\,p$ and $\det M_\bq=\nu^2(\gamma^2+B^2)^2\,r$, with $p$ and $r$ as defined in the main text. The quadratic then becomes
\begin{equation}
\lambda^2 + 2\nu p\,\lambda + \nu^2 r = 0
\quad\Longrightarrow\quad
\lambda_\pm = -\nu\Bigl[p\pm\sqrt{p^2-r}\Bigr],
\end{equation}
which coincides with Eq.~\eqref{diffusiveeigenvalues}, with $\nu$, $p$ and $r$ given by Eq.~\eqref{eq:nupr}. The exceptional points, $p^2=r$, coincide with Eq.~\eqref{exceptionalline}, which has been used in Fig.~\ref{fig:phasediagram}(a). The real and imaginary parts of $\lambda_\pm$ give the decay rate and the oscillation frequency of Eqs.~\eqref{eq:tau_Omega} and \eqref{eq:tau_explicit}, which have been used in Fig.~\ref{fig:dynamics}(e).

\section{Eigenvalue Analysis at the $\bK$ Point}
\label{Kpoint}

At the corners of the Brillouin zone, the characteristic polynomial~\eqref{eqquartogrado_final} factorizes as
\begin{equation}
P(\lambda)=\Bigl(\lambda^2 +\gamma \lambda +\omega_\bK^2 +i \alpha_\bK^2 - iB \lambda\Bigr) \Bigl(\lambda^2 +\gamma \lambda +\omega_\bK^2 -i \alpha_\bK^2 +iB \lambda \Bigr).
\label{fulldispersionequationlast0}
\end{equation}
The solutions are, with $\lambda_R$ and $\lambda_I$ the real and imaginary parts of the eigenvalue $\lambda$, given by:
\begin{subequations}
\begin{align}
\lambda^{(1,2)}_R &= -\frac{\gamma}{2} + \frac{1}{\sqrt{2}} \sqrt{ \sqrt{\mathcal{A}^2 + \mathcal{C}^2} + \mathcal{A} }, &
\lambda^{(1,2)}_I &= \pm\Bigl( \frac{B}{2} - \frac{1}{\sqrt{2}} \sqrt{ \sqrt{\mathcal{A}^2 + \mathcal{C}^2} - \mathcal{A} }\Bigr), \label{undampedmode}\\
\lambda^{(3,4)}_R &= -\frac{\gamma}{2} -\frac{1}{\sqrt{2}} \sqrt{ \sqrt{\mathcal{A}^2 + \mathcal{C}^2} + \mathcal{A} }, &
\lambda^{(3,4)}_I &= \pm\Bigl( \frac{B}{2} +\frac{1}{\sqrt{2}} \sqrt{ \sqrt{\mathcal{A}^2 + \mathcal{C}^2} - \mathcal{A} }\Bigr), \label{dampedmode}
\end{align}
\end{subequations}
with
\begin{equation}
\mathcal{A}=\frac{\gamma^2-B^2}{4}-\omega_\bK^2, \qquad \mathcal{C}=\frac{\gamma B}{2}+\alpha_\bK^2.
\end{equation}
At the Brillouin-zone corners, $\lambda^{(1,2)}_R$ vanishes if
\begin{equation}
(\alpha_\bK^2)^2 + \gamma B \alpha_\bK^2 = \gamma^2 \omega_\bK^2,
\label{Kinstability}
\end{equation}
with $\alpha_\bK^2$ and $\omega_\bK^2$ from Eqs.~\eqref{mk2}-\eqref{mk3}, giving a pair of undamped modes, as shown by Eq.~\eqref{undampedmode}. Equation~\eqref{Kinstability} coincides with Eq.~\eqref{Kinstability0} of the main text, and Eqs.~\eqref{undampedmode}--\eqref{dampedmode} have been used in Fig.~\ref{fig:phasediagram}(c).
The full spectrum along the high-symmetry path of the Brillouin zone is shown in Fig.~\ref{fig:spectrum}.

\section{Equal-Time Correlations with Magnetic Field}
\label{Lyapunovequations}

The Langevin dynamics in matrix form, using displacement and velocity variables, is
\begin{equation}
\frac{d}{dt }\begin{pmatrix} \tilde{u}_\bq^x \\ \tilde{v}_\bq^x \\ \tilde{u}_\bq^y \\ \tilde{v}_\bq^y \end{pmatrix}
=-\begin{pmatrix} 0 & -1 & 0 &0 \\ -M_{xx}& \gamma & -M_{xy}&B\\ 0 & 0& 0 &-1 \\ -M_{yx} &-B & -M_{yy}& \gamma \end{pmatrix}
\begin{pmatrix} \tilde{u}_\bq^x \\ \tilde{v}_\bq^x \\ \tilde{u}_\bq^y \\ \tilde{v}_\bq^y \end{pmatrix}
+\sqrt{\frac{2\gamma T}{m}}
\begin{pmatrix} 0 \\ \tilde\xi_\bq^x \\ 0 \\ \tilde\xi_\bq^y \end{pmatrix},
\end{equation}
with $\langle \tilde\xi^\alpha_\bq(t) \tilde\xi^\beta_{\bq'}(t') \rangle= \delta(t-t')\,\delta^{\alpha\beta}\, \delta_{\bq+\bq',0}$. For convenience, we define $\omega_\bq^2=-\Tr(M)/2$ and $\alpha_\bq^2=(M_\bq^{xy}-M_\bq^{yx})/2$.
By introducing the vector $X=(\tilde{u}_\bq^x,\tilde{v}_\bq^x,\tilde{u}_\bq^y,\tilde{v}_\bq^y)$, we can express the dynamics in a compact form $\dot X_i=-A_{ij}X_j+\Phi_{ij}\xi_j$, with $\langle\xi_i(t)\xi_j(t')\rangle=\delta(t-t')\delta_{ij}$ and $\Phi_{22}=\Phi_{44}=\sqrt{2\gamma T/m}$ (zero otherwise). In this way, we have $\sum_k\Phi_{ik}\Phi_{kj}=2D_{ij}$. The equal-time covariance follows from the Lyapunov equation
\begin{equation}
D_{ij}=\frac{1}{2}\Bigl( A_{ik} C_{kj}+C_{ik} (A^T)_{kj}\Bigr).
\label{eq:lyap4x4}
\end{equation}
The exact solution of Eq.~\eqref{eq:lyap4x4} has been used in Figs.~\ref{fig:velocity}(a)--(h) and \ref{fig:structure}(b)--(e). For $\bq=\bK$, the equal-time correlations are given by Eqs.~\eqref{Eq:variance1}--\eqref{Eq:variance3}. For generic anisotropic $\bq$ the Lyapunov equation above can in principle be solved directly (e.g.\ numerically, or symbolically via computer algebra), but the resulting closed-form expressions are algebraically cumbersome and are not reported here; this is why, as stated in Sec.~\ref{Staticproperties}, we restrict the analytical discussion to the two tractable limits $\bq\to0$ and $\bq\simeq\bK$. In the isotropic special case, $M_\bq^{xx}=M_\bq^{yy}$ and $M_\bq^{xy}=-M_\bq^{yx}$, the anisotropic cross-term vanishes identically for any $\bq$,
\begin{equation}
\langle \tilde{v}_\bq^x \tilde{v}_{-\bq}^x - \tilde{v}_\bq^y \tilde{v}_{-\bq}^y \rangle = 0,
\end{equation}
and the velocity sum reduces to the same functional form found at the $\bK$ point, given by
\begin{equation}
\langle \tilde{v}_\bq^x \tilde{v}_{-\bq}^x + \tilde{v}_\bq^y \tilde{v}_{-\bq}^y \rangle = \frac{2T}{m}\,\frac{\gamma^2\omega_\bq^2-B\gamma\alpha_\bq^2}{\gamma^2\omega_\bq^2-B\gamma\alpha_\bq^2-(\alpha_\bq^2)^2},
\end{equation}
with $\omega_\bq^2=-\Tr(M_\bq)/2$, $\alpha_\bq^2=(M_\bq^{xy}-M_\bq^{yx})/2$ evaluated at the wavevector of interest (both verified directly against the exact solution of the $4\times4$ Lyapunov equation).

\noindent
\vskip8pt
\noindent
\textbf{Long-wavelength limit: adiabatic elimination of the velocities --} For $\gamma\gg\omega_\bq$, the velocities relax on the fast time scale $1/\gamma$ and can be eliminated by setting $\dot{\tilde{\mathbf v}}_\bq\simeq0$ in Eq.~\eqref{eq:fourier_space_eq}. Writing the friction and Lorentz terms as $\mathbf{\Gamma}\,\tilde{\mathbf v}_\bq$, with
\begin{equation}
\mathbf{\Gamma}=\begin{pmatrix}\gamma & B\\ -B & \gamma\end{pmatrix},
\end{equation}
one obtains $\tilde{\mathbf v}_\bq=\mathbf{\Gamma}^{-1}\bigl[\mathbf M_\bq\tilde{\mathbf u}_\bq+\sqrt{2\gamma T/m}\,\tilde{\boldsymbol\xi}_\bq\bigr]$, so that the displacements obey the first-order dynamics $\dot{\tilde{\mathbf u}}_\bq=\mathbf{\Gamma}^{-1}\mathbf M_\bq\tilde{\mathbf u}_\bq+\sqrt{2\gamma T/m}\,\mathbf{\Gamma}^{-1}\tilde{\boldsymbol\xi}_\bq$. The steady-state displacement covariance, $\mathbf\Sigma_\bq=\langle\tilde{\mathbf u}_\bq\tilde{\mathbf u}_{-\bq}^T\rangle$, satisfies the $2\times2$ Lyapunov equation $\mathbf{\Gamma}^{-1}\mathbf M_\bq\mathbf\Sigma_\bq+\mathbf\Sigma_\bq\mathbf M_\bq^T\mathbf{\Gamma}^{-T}+\frac{2\gamma T}{m}\mathbf{\Gamma}^{-1}\mathbf{\Gamma}^{-T}=0$. Multiplying from the left by $\mathbf\Gamma$ and from the right by $\mathbf\Gamma^T$, it takes the simpler form
\begin{equation}
\mathbf M_\bq\mathbf\Sigma_\bq\mathbf{\Gamma}^T+\mathbf{\Gamma}\mathbf\Sigma_\bq\mathbf M_\bq^T=-\frac{2\gamma T}{m}\,\mathbf{I}\,.
\label{eq:lyap2x2}
\end{equation}
For $\bq=q\hat{\mathbf x}$, the long-wavelength matrix elements of Eqs.~\eqref{M52}--\eqref{M55} make the three independent components of Eq.~\eqref{eq:lyap2x2} a linear system for $\Sigma^{xx}_\bq$, $\Sigma^{yy}_\bq$ and $\Sigma^{xy}_\bq$, whose solution is
\begin{equation}
\Sigma^{xx}_\bq=\frac{8}{9}\,\frac{T}{C_0q^2}\,\frac{1}{1-\frac{B}{\gamma}\frac{C_1}{C_0}},\qquad
\Sigma^{yy}_\bq=3\,\Sigma^{xx}_\bq,\qquad
\Sigma^{xy}_\bq=0.
\end{equation}
For this choice of $\bq$, $\Sigma^{xx}_\bq$ and $\Sigma^{yy}_\bq$ are the longitudinal and transverse correlations, $\Sigma_{LL}(\bq)$ and $\Sigma_{TT}(\bq)$, and coincide with Eqs.~\eqref{longitudinalcorrelation} and \eqref{transversecorrelation}. Since the triangular lattice is elastically isotropic at long wavelengths, the same expressions hold for any direction of $\bq$, as we have checked against the exact solution of the $4\times4$ Lyapunov equation.

\vskip8pt
\noindent
\textbf{Brillouin-zone corner --} At $\bq=\bK$, the dynamical matrix takes the isotropic form $\mathbf M_\bK=-\omega_\bK^2\mathbf I+\alpha_\bK^2\boldsymbol\epsilon$, Eqs.~\eqref{mk2}--\eqref{mk3}, and the dynamics decouples in the complex variables $z_u=\tilde u^x_\bK+i\tilde u^y_\bK$ and $z_v=\tilde v^x_\bK+i\tilde v^y_\bK$, which obey
\begin{equation}
\dot z_u=z_v,\qquad
\dot z_v=-\bigl(\omega_\bK^2+i\alpha_\bK^2\bigr)z_u-(\gamma-iB)\,z_v+\sqrt{\frac{2\gamma T}{m}}\,\eta,
\end{equation}
where $\eta=\tilde\xi^x_\bK+i\tilde\xi^y_\bK$ is a complex white noise with $\langle\eta(t)\eta^*(t')\rangle=2\delta(t-t')$. The required correlations are $C_{uu}=\langle|z_u|^2\rangle$, $C_{vv}=\langle|z_v|^2\rangle$ and $\langle z_u^*z_v\rangle$, whose imaginary part is $\langle\tilde u^x\tilde v^y-\tilde u^y\tilde v^x\rangle\equiv y$. Imposing stationarity, $d\langle|z_u|^2\rangle/dt=0$ gives ${\rm Re}\langle z_u^*z_v\rangle=0$, so that $\langle z_u^*z_v\rangle=iy$, while $d\langle z_u^*z_v\rangle/dt=0$ and $d\langle|z_v|^2\rangle/dt=0$ give
\begin{equation}
C_{vv}-\omega_\bK^2C_{uu}-By=0,\qquad
\alpha_\bK^2C_{uu}+\gamma y=0,\qquad
\gamma C_{vv}+\alpha_\bK^2y=\frac{2\gamma T}{m}.
\end{equation}
Solving this linear system yields
\begin{equation}
C_{uu}=\frac{2T}{m}\frac{\gamma^2}{\Delta_\bK},\qquad
C_{vv}=\frac{2T}{m}\frac{\gamma^2\omega_\bK^2-B\gamma\alpha_\bK^2}{\Delta_\bK},\qquad
y=-\frac{2T}{m}\frac{\gamma\alpha_\bK^2}{\Delta_\bK},
\end{equation}
with $\Delta_\bK=\gamma^2\omega_\bK^2-B\gamma\alpha_\bK^2-(\alpha_\bK^2)^2$, which coincide with Eqs.~\eqref{Eq:variance1}--\eqref{Eq:variance3} of the main text. In particular, $y$ has the sign opposite to $C_1$, so that the contribution of the modes at $\bK$ to the angular momentum is negative for $C_1>0$.

\vskip8pt
\noindent
\textbf{Numerical evaluation at arbitrary wavevector --} For generic $\bq$, the Lyapunov equation is a linear system of $16$ equations for the elements of the covariance matrix $C_\bq$, which we solve exactly for each allowed wavevector of an $L\times L$ lattice with periodic boundary conditions, $\bq=(l_1\mathbf G_1+l_2\mathbf G_2)/L$, excluding the Goldstone mode $\bq=0$. Real-space correlations are obtained by inverse discrete Fourier transform, e.g.
\begin{equation}
\langle\mathbf v_\bn\cdot\mathbf v_{\mathbf 0}\rangle=\frac1N\sum_{\bq\neq0}\langle\tilde{\mathbf v}_\bq\cdot\tilde{\mathbf v}_{-\bq}\rangle\,e^{i\bq\cdot\bR^0_\bn},
\label{eq:realspace}
\end{equation}
and single-particle averages, such as $T_{\rm kin}$ and $L$, by setting $\bn=\mathbf 0$. 
Equation~\eqref{eq:realspace} has been used in Fig.~\ref{fig:velocity}(e), and its single-particle limit in Fig.~\ref{fig:velocity}(g),(h). In Fig.~\ref{fig:velocity} we used a lattice of $120\times120$ sites for the real-space correlations and of $48\times48$ sites for $T_{\rm kin}$ and the angular momentum. Close to the instability at $\bK$, the velocity spectrum takes the Ornstein--Zernike form $\langle\tilde{\mathbf v}_\bq\cdot\tilde{\mathbf v}_{-\bq}\rangle\simeq A_\xi/[1+\xi^2|\bq-\bK|^2]$, with an amplitude $A_\xi\propto\xi^2$, since the peak height diverges as the inverse distance from the instability. Summing over $\bq$ in two dimensions, the contribution of the peak to single-particle averages scales as $\int d^2\delta q\,\xi^2/(1+\xi^2\delta q^2)\propto\ln(\xi/a)$, which explains the logarithmic divergence of $T_{\rm kin}$ and of the angular momentum. Time-dependent correlations are obtained as
\begin{equation}
\langle X_\bq(t)X^\dagger_\bq(0)\rangle=e^{-At}C_\bq\,,\qquad t\geq0,
\label{eq:timecorr}
\end{equation}
and spectra as $C_\bq(\omega)=(-i\omega+A)^{-1}2D\,(i\omega+A^{T})^{-1}$, equivalent to Eq.~\eqref{eq:CABresult}, with $A$ and $D$ defined above. Equation~\eqref{eq:timecorr} has been used to compute the intermediate scattering function of Fig.~\ref{fig:dynamics}(d). In particular, the single-particle velocity autocorrelation is obtained by summing the velocity components of Eq.~\eqref{eq:timecorr} over all the allowed wavevectors,
\begin{equation}
\langle\mathbf v_\bn(t)\cdot\mathbf v_\bn(0)\rangle=\frac1N\sum_{\bq\neq0}\langle\tilde{\mathbf v}_\bq(t)\cdot\tilde{\mathbf v}_{-\bq}(0)\rangle\,.
\label{eq:vacf}
\end{equation}
Equation~\eqref{eq:vacf} has been used in Fig.~\ref{fig:dynamics}(f).
The correlation length is defined from the curvature of the peak at $\bK$,
\begin{equation}
\xi^2=-\frac{\partial_q^2\langle\tilde{\mathbf v}_\bq\cdot\tilde{\mathbf v}_{-\bq}\rangle}{2\langle\tilde{\mathbf v}_\bK\cdot\tilde{\mathbf v}_{-\bK}\rangle}\bigg|_{\bq=\bK},
\label{eq:xi}
\end{equation}
evaluated along $q_x$. 
Equation~\eqref{eq:xi} has been used in Fig.~\ref{fig:velocity}(f).

\section{Derivation of the Static Structure Factor}
\label{staticSk}

The microscopic density of the crystal is
\begin{equation}
\rho(\mathbf r) = \sum_n \delta\!\left( \mathbf r-\mathbf R^0_n-\mathbf u_n \right),
\end{equation}
where $\mathbf R^0_n$ denotes the equilibrium lattice position of particle $n$ and $\mathbf u_n$ its displacement. The Fourier transform of the density is
\begin{equation}
\rho(\mathbf k) = \sum_n e^{-i\mathbf k\cdot(\mathbf R^0_n+\mathbf u_n)},
\end{equation}
and the static structure factor is
\begin{equation}
S(\mathbf k) = \frac{1}{N} \left\langle \rho(\mathbf k)\rho(-\mathbf k) \right\rangle.
\label{eq:Sk_master}
\end{equation}
The two limits considered below correspond to two physically distinct, and separately linearized, regimes of Eq.~\eqref{eq:Sk_master}. Its one-phonon expansion, $S(\bk)=k_\alpha k_\beta\langle\tilde u^\alpha_\bk\tilde u^\beta_{-\bk}\rangle$, has been used in Fig.~\ref{fig:structure}(b)--(e).

\subsection{Diffuse scattering near a Bragg peak}
\label{app:bragg}

To study the diffuse scattering surrounding a Bragg peak, $\bk=\mathbf G$, we write $\mathbf k=\mathbf G+\bq$ with $\mathbf G\neq0$ a reciprocal lattice vector and $|\bq|\ll|\mathbf G|$. Within the harmonic approximation,
\begin{equation}
e^{-i(\mathbf G+\bq)\cdot\mathbf u_n} \simeq 1 - i(\mathbf G+\bq)\cdot\mathbf u_n \simeq 1-i\,\mathbf G\cdot\mathbf u_n,
\end{equation}
where in the last step we used $|\bq|\ll|\mathbf G|$ to drop $\bq\cdot\mathbf u_n$ relative to $\mathbf G\cdot\mathbf u_n$. This is legitimate because $\mathbf G$, though finite, is compared with $\bq\to0$, and not because $\mathbf G\cdot\mathbf u_n$ is assumed to be small. The full $\mathbf G+\bq$ is instead retained in the phase $e^{-i(\mathbf G+\bq)\cdot\mathbf R_n}$, since $\bq$ alone selects the wavevector of the displacement correlations probed near the Bragg spot. This gives
\begin{equation}
\delta\rho(\mathbf G+\bq) = -i \sum_n (\mathbf G\cdot\mathbf u_n)\, e^{-i(\mathbf G+\bq)\cdot\mathbf R_n},
\end{equation}
so that the diffuse contribution to the structure factor is
\begin{align}
S(\mathbf G+\bq)
&= \frac{1}{N} \left\langle \delta\rho(\mathbf G+\bq)\, \delta\rho(-\mathbf G-\bq) \right\rangle \nonumber\\
&= G_\alpha G_\beta \frac{1}{N} \sum_{nm} e^{-i\bq\cdot(\mathbf R_n-\mathbf R_m)} \left\langle u_n^\alpha u_m^\beta \right\rangle.
\end{align}
Introducing the Fourier-space displacement correlations $\langle \tilde{u}^\alpha_{\bq} \tilde{u}^\beta_{-\bq}\rangle = \frac{1}{N}\sum_{nm} e^{-i\bq\cdot(\mathbf R_n-\mathbf R_m)} \langle u_n^\alpha u_m^\beta\rangle$, we can express $S(\mathbf G+\bq)$ as
\begin{equation}
S(\mathbf G+\bq) = G_\alpha G_\beta \left\langle \tilde{u}^\alpha_{\bq} \tilde{u}^\beta_{-\bq} \right\rangle.
\label{eq:SGq_general}
\end{equation}
The displacement correlation tensor decomposes into longitudinal and transverse parts,
\begin{equation}
\left\langle \tilde{u}^\alpha_{\bq} \tilde{u}^\beta_{-\bq} \right\rangle = \Sigma_{LL}(\bq)\, P_L^{\alpha\beta} + \Sigma_{TT}(\bq)\, P_T^{\alpha\beta},
\qquad
P_L^{\alpha\beta} = \hat q^\alpha\hat q^\beta,
\qquad
P_T^{\alpha\beta} = \delta^{\alpha\beta} - \hat q^\alpha\hat q^\beta,
\end{equation}
so that Eq.~\eqref{eq:SGq_general} becomes
\begin{equation}
S(\mathbf G+\bq) = \Sigma_{LL}(\bq)\,(\mathbf G\cdot\hat{\bq})^2 + \Sigma_{TT}(\bq)\left[ G^2 - (\mathbf G\cdot\hat{\bq})^2 \right].
\label{eq:SGq_decomposed}
\end{equation}
Using the long-wavelength correlations (Sec.~\ref{Staticproperties}),
\begin{align}
\Sigma_{LL}(\bq) &= \frac{8}{9} \frac{T}{C_0} \frac{1}{q^2} \frac{1}{1-\frac{B}{\gamma}\frac{C_1}{C_0}},
&
\Sigma_{TT}(\bq) &= \frac{8}{3} \frac{T}{C_0} \frac{1}{q^2} \frac{1}{1-\frac{B}{\gamma}\frac{C_1}{C_0}},
\label{eq:sigmaLLTT_recall}
\end{align}
Eq.~\eqref{eq:SGq_decomposed} gives
\begin{equation}
S(\mathbf G+\bq) = \frac{8T}{C_0} \frac{1}{q^2} \frac{1}{1-\frac{B}{\gamma}\frac{C_1}{C_0}} \left[ \frac{1}{9}(\mathbf G\cdot\hat{\bq})^2 + \frac{1}{3}\Bigl( G^2-(\mathbf G\cdot\hat{\bq})^2 \Bigr) \right].
\label{SGQnew}
\end{equation}
Taking, for concreteness, one of the smallest reciprocal lattice vectors, $\mathbf G=\frac{2\pi}{a}(1,1/\sqrt3)$, and writing $\cos\theta = \hat{\mathbf G}\cdot\hat{\bq}$ (so that $\mathbf G\cdot\hat{\bq}=G\cos\theta$), this becomes
\begin{equation}
S(\mathbf G+\bq) = \frac{8G^2T}{3C_0}\,\frac{1}{q^2}\,\frac{1}{1-\frac{B}{\gamma}\frac{C_1}{C_0}}\left[1-\frac{2}{3}\cos^2\theta\right].
\label{eq:SGq_final_app}
\end{equation}
This expression coincides with Eq.~\eqref{eq:SGq_final} of the main text.
Equation~\eqref{eq:SGq_final_app} is the leading-order (one-phonon, harmonic) truncation of the fully Debye--Waller-resummed structure factor, $S(\mathbf G+\bq)\propto e^{-\langle(\mathbf G\cdot\mathbf u)^2\rangle}\times(\text{one-phonon term}+\cdots)$, valid for $\bq$ sufficiently close to, but not asymptotically at, the Bragg point. In strict two dimensions, $\langle u^2\rangle=\int d^2q\,\langle|\tilde{u}_\bq|^2\rangle$ diverges logarithmically at small $q$ -- the Mermin--Wagner statement that a 2D crystal possesses only quasi-long-range positional order -- so that resumming the full Debye--Waller factor turns the true Bragg peak into an algebraic singularity, $S(\mathbf G+\bq)\sim q^{-2+\eta(T)}$, with a temperature-dependent exponent $\eta$, rather than the bare $1/q^2$ found here. Since the magnetic field and odd elasticity already enhance or suppress the displacement fluctuations entering $\langle u^2\rangle$, a fully resummed treatment would be expected to yield a $B$- and $C_1$-tunable anomalous exponent $\eta$ -- odd elasticity and the Lorentz force acting as external control parameters on the degree of positional order of the two-dimensional lattice. We leave this resummation to future work.

\subsection{Forward-scattering (compressibility) limit}
\label{app:compressibility}

The second physically distinct limit is $\bk\to0$ directly, with no reciprocal lattice vector involved. Here $\bk$ itself is the small parameter, and the relevant linearization is a mass-conservation identity rather than a harmonic expansion around a Bragg point: for small displacements, $\delta\rho(\mathbf r,t)=-\rho_0\,\nabla\cdot\mathbf u(\mathbf r,t)$, independently of the underlying force law. In Fourier space, $\delta\rho_\bq=-i\rho_0\,\bq\cdot\tilde{\mathbf u}_\bq$, so that only the \emph{longitudinal} component of the displacement -- the projection along $\bq$ -- contributes; the transverse (shear) component is divergence-free and drops out identically. The static structure factor at small $\bq$ is therefore
\begin{equation}
S(\bq) \equiv \frac{\langle \delta\rho_\bq\,\delta\rho_{-\bq}\rangle}{\rho_0^2 N} = q^2\,\Sigma_{LL}(\bq),
\label{eq:Sq_def}
\end{equation}
with no contribution from $\Sigma_{TT}(\bq)$ at this order, despite the latter being three times larger. Substituting Eq.~\eqref{eq:sigmaLLTT_recall}, the factor $q^2$ cancels the $1/q^2$ divergence of $\Sigma_{LL}$ exactly, leaving a finite $q\to0$ limit,
\begin{equation}
S(0) = \frac{8}{9}\,\frac{T}{C_0}\,\frac{1}{1-\frac{B}{\gamma}\frac{C_1}{C_0}}.
\label{eq:S0_app}
\end{equation}
This expression coincides with Eq.~\eqref{eq:S0} of the main text and has been used in Figs.~\ref{fig:phasediagram}(d) and \ref{fig:structure}(a). In the absence of odd elasticity or magnetic field, this reduces to the familiar equilibrium compressibility sum rule, $S(0)\propto T/C_0$, set by the ordinary bulk modulus alone. The combined action of the magnetic field and odd elasticity renormalizes the forward-scattering intensity by the same factor that renormalizes the long-wavelength displacement fluctuations (Sec.~\ref{Staticproperties}): a single control parameter, $BC_1/(\gamma C_0)$, tunes both quantities simultaneously, and either could in principle be used to measure the other. As $BC_1/(\gamma C_0)\to1^-$, $S(0)$ diverges: a long-wavelength, dynamical (compressibility-type) instability, distinct from the finite-wavevector instability of Sec.~\ref{sec:Kpoint_instability}. 
For $w>0$, this divergence lies beyond the onset of the finite-wavevector instability at $\bK$, while the two coincide in the overdamped limit $w\to0$ (Sec.~\ref{Staticproperties}).

\section{Correlation functions in the frequency domain}
\label{app:frequency}

Assuming periodic (toroidal) boundary conditions, we use a Fourier representation in space and time,
\begin{equation}
\tilde{\mathbf u}_\bq(\omega) = \frac{1}{\sqrt{N}} \int_{-\infty}^{\infty} dt\, e^{i\omega t} \sum_\bn \rbu_\bn(t)\, e^{-i \bq \cdot \bR^0_\bn}.
\end{equation}
The Langevin equation \eqref{ALangevin1} takes the compact form
\begin{equation}
L_\bq^{\alpha\beta}(\omega) \tilde{u}_\bq^\beta(\omega) = \sqrt{\frac{2\gamma T}{m}} \,\tilde \xi_\bq^\alpha(\omega),
\label{eq:stochasticequation}
\end{equation}
with
\begin{equation}
L_\bq^{\alpha\beta}(\omega) = -(\omega^2+i\gamma\omega)\delta^{\alpha\beta} -M_\bq^{\alpha\beta} -i\omega B\,\epsilon^{\alpha\beta}.
\label{linearoperatorL}
\end{equation}
The dynamical matrix $M_\bq$ contains all the interaction and geometric information (Appendix \ref{matrixelements}), and the Green's function is $G_{\mathbf q}(\omega)=L_{\mathbf q}^{-1}(\omega)$. Since the noise is Gaussian and uncorrelated, $\langle \tilde\xi^\alpha_\bq(\omega) (\tilde\xi^\beta_\bq(\omega))^*\rangle=\delta^{\alpha\beta}$, we obtain
\begin{equation}
C_\bq^{\alpha\beta}(\omega) \equiv \langle \tilde{u}_\bq^\alpha(\omega)(\tilde{u}_{-\bq}^\beta(\omega))^*\rangle = \frac{2\gamma T}{m}\, G_\bq^{\alpha\mu}(\omega)(G_\bq^{\beta\mu}(\omega))^*.
\label{eq:CABresult}
\end{equation}
Equation~\eqref{eq:CABresult} has been used in Fig.~\ref{fig:dynamics}(a)--(c).

\section{Two-time correlations in the overdamped approximation}
\label{app:twotime}
In this Appendix we derive the two-time correlations of the long-wavelength modes and the intermediate scattering function used in Sec.~\ref{sec:isf}. Using Eq.~\eqref{eq:CABresult}, for $\bq$ directed along the $x$ axis, the two-time longitudinal and transverse correlations are
\begin{subequations}
\begin{align}
C^{LL}_\bq(\omega) &= \langle \tilde{u}^x_\bq(\omega) \tilde{u}^x_{-\bq}(-\omega) \rangle = 2 \gamma\frac{T}{m}\frac{ (M_\bq^{yy})^2+(M_\bq^{xy})^2+\omega^2(\gamma^2+B^2) }{ D_\bq(\omega) }, \label{Vitelliuxux} \\
C^{TT}_\bq(\omega) &= \langle \tilde{u}^y_\bq(\omega) \tilde{u}^y_{-\bq}(-\omega) \rangle = 2 \gamma\frac{T}{m}\frac{(M_\bq^{xx})^2+(M_\bq^{yx})^2+\omega^2(\gamma^2+B^2) }{ D_\bq(\omega) },\label{Vitelliuuyuy}
\end{align}
\end{subequations}
with
\begin{equation}
D_\bq(\omega) = \Bigl[ \bigl( (\gamma^2+B^2)\omega^2 -\det(M_\bq)\bigr)^2 +\omega^2 \bigl(\gamma \Tr(M_\bq)+B(M_\bq^{xy}-M_\bq^{yx})\bigr)^2\Bigr],
\end{equation}
which factorizes in terms of the diffusive eigenvalues $\lambda_\pm$, Eq.~\eqref{diffusiveeigenvalues}:
\begin{equation}
D_\bq(\omega)=(\gamma^2+B^2)^2 (\omega-i\lambda_{+}) (\omega-i\lambda_{-}) (\omega+i\lambda^*_{+}) (\omega+i\lambda^*_{-}).
\label{denominatordynamicvv3}
\end{equation}
Using Eqs.~\eqref{M52}-\eqref{M55},
\begin{subequations}
\begin{align}
C_\bq^{LL}(t) &= \frac{2\gamma T}{m(\gamma^2+B^2)^2}\left[ \frac{9 q^4} {64 m^2} (C_0^2+C_1^2)\right] P_0(t)+\frac{2\gamma T}{m(\gamma^2+B^2) } P_2(t), \\
C_\bq^{TT}(t) &= \frac{2\gamma T}{m(\gamma^2+B^2)^2}\left[ \frac{81 q^4} {64 m^2} (C_0^2+C_1^2)\right] P_0(t) + \frac{2\gamma T}{m(\gamma^2+B^2)} P_2(t),
\end{align}
\end{subequations}
where
\begin{subequations}
\begin{align}
P_0(t) &= \int_{-\infty}^{\infty} \frac{d\omega}{2\pi}\, \frac{e^{-i\omega t}} {(\omega-i\lambda_{+}) (\omega-i\lambda_{-}) (\omega+i\lambda^*_{+}) (\omega+i\lambda^*_{-})}, \\
P_2(t) &= \int_{-\infty}^{\infty} \frac{d\omega}{2\pi}\, \frac{\omega^2 e^{-i\omega t}}{(\omega-i\lambda_{+}) (\omega-i\lambda_{-}) (\omega+i\lambda^*_{+}) (\omega+i\lambda^*_{-})}.
\end{align}
\end{subequations}
The time-dependent correlations determine the intermediate scattering function. Repeating the linearization of Appendix~\ref{staticSk} at unequal times -- $\delta\rho(\mathbf k,t)\simeq -i\sum_n(\mathbf k\cdot\mathbf u_n(t))\,e^{-i\mathbf k\cdot\mathbf R_n}$, valid near a Bragg point $\mathbf k=\mathbf G+\bq$, $|\bq|\ll|\mathbf G|$, or directly at small $\mathbf k=\bq$ for the forward-scattering limit, and using the same longitudinal/transverse decomposition of the \emph{time-dependent} displacement correlation tensor,
\begin{equation}
\left\langle \tilde{u}^\alpha_{\bq}(t)\, \tilde{u}^\beta_{-\bq}(0) \right\rangle = C^{LL}_\bq(t)\, P_L^{\alpha\beta} + C^{TT}_\bq(t)\, P_T^{\alpha\beta},
\end{equation}
gives, in exact analogy with Eqs.~\eqref{eq:SGq_decomposed} and \eqref{eq:Sq_def},
\begin{subequations}
\begin{align}
F(\mathbf G+\bq,t) &= C^{LL}_\bq(t)\,(\mathbf G\cdot\hat{\bq})^2 + C^{TT}_\bq(t)\left[ G^2 - (\mathbf G\cdot\hat{\bq})^2 \right], \label{eq:FGqt}\\
F(\bq,t) &= q^2\, C^{LL}_\bq(t), \label{eq:Fqt}
\end{align}
\end{subequations}
the dynamic generalizations of Eqs.~\eqref{eq:SGq_decomposed} and \eqref{eq:Sq_def} respectively; both reduce to the corresponding static result at $t=0$, since $C^{LL}_\bq(0)=\Sigma_{LL}(\bq)$, $C^{TT}_\bq(0)=\Sigma_{TT}(\bq)$. As in the static case, only the longitudinal correlation enters the forward-scattering limit, Eq.~\eqref{eq:Fqt}; the transverse (shear) relaxation is invisible to $F(\bq,t)$ at small $\bq$ despite dominating the equal-time variance.

We now evaluate the integrals $P_0(t)$ and $P_2(t)$. Closing the integration contour in the lower half plane for $t>0$, the integrals are given by the residues at the poles $\omega=i\lambda_\pm$. For $p^2>r$, the eigenvalues $\lambda_\pm$ are real and
\begin{subequations}
\begin{align}
P_0(t) &= \frac{1}{8\nu^2 p\sqrt{p^2-r} }\Bigl[ \frac{1}{\lambda_{+}} e^{\lambda_{+} |t|} -\frac{1}{\lambda_{-}} e^{\lambda_{-} |t|} \Bigr], \\
P_2(t) &= \frac{1}{8 \nu^2 p\sqrt{p^2-r} }\Bigl[-\lambda_{+} e^{\lambda_{+} |t|} + \lambda_{-} e^{\lambda_{-} |t|}\Bigr].
\end{align}
\end{subequations}
For $r>p^2$, the eigenvalues are complex conjugate, $\lambda_\pm=-\nu p\mp i\nu\sqrt{r-p^2}$, and
\begin{subequations}
\begin{align}
P_0(t) &= \frac{e^{-\nu p|t|}} {4\nu^3rp\sqrt{r-p^2}} \left[ \sqrt{r-p^2}\, \cos\!\bigl(\nu\sqrt{r-p^2}|t|\bigr) + p\, \sin\!\bigl(\nu\sqrt{r-p^2}|t|\bigr) \right], \\
P_2(t) &= -\frac{d^2 P_0}{dt^2}= \frac{e^{-\nu p|t|}} {4\nu p\sqrt{r-p^2}} \left[ \sqrt{r-p^2}\, \cos\!\bigl(\nu\sqrt{r-p^2}\,|t|\bigr)-p\,\sin\!\bigl(\nu\sqrt{r-p^2}\,|t|\bigr) \right].
\end{align}
\end{subequations}
At $t=0$, both regimes give $P_0(0)=1/(4\nu^3rp)$ and $P_2(0)=1/(4\nu p)$, so that the correlations of Sec.~\ref{Timedependentcorrelations} reduce to
\begin{subequations}
\begin{align}
C_\bq^{LL}(0) &= \frac{8}{9}\,\frac{T}{C_0 q^2}\, \frac{1}{1 - \frac{B}{\gamma}\frac{C_1}{C_0}}, \\
C_\bq^{TT}(0) &= \frac{8}{3}\,\frac{T}{C_0 q^2}\, \frac{1}{1 - \frac{B}{\gamma}\frac{C_1}{C_0}}.
\end{align}
\end{subequations}
which coincide with Eqs.~\eqref{longitudinalcorrelation} and \eqref{transversecorrelation}, with the expected factor 3 between transverse and longitudinal correlations.

\vskip8pt
\noindent
\textbf{Intermediate scattering function --} Using $\nu=\tfrac34 q^2/(\gamma^2+B^2)$ and the definition of $r$, the $q^4$ prefactor in $C^{LL}_\bq(t)$ can be written as $\frac{9q^4}{64m^2}(C_0^2+C_1^2)=\frac13(\gamma^2+B^2)^2\nu^2 r$, while $q^2=\frac43(\gamma^2+B^2)\,\nu$. Inserting $C^{LL}_\bq(t)$ into Eq.~\eqref{eq:Fqt} then gives
\begin{equation}
F(\bq,t)=\frac{8\gamma T}{3m}\,\nu\left[\frac{\nu^2 r}{3}\,P_0(t)+P_2(t)\right],
\end{equation}
with no explicit $q$-dependence beyond that contained in $\nu\propto q^2$. In the oscillatory regime, the two terms in the brackets give $\frac{e^{-\nu p|t|}}{12\,p\sqrt{r-p^2}}\bigl[\sqrt{r-p^2}\cos(\Omega|t|)+p\sin(\Omega|t|)\bigr]$ and $\frac{e^{-\nu p|t|}}{4\,p\sqrt{r-p^2}}\bigl[\sqrt{r-p^2}\cos(\Omega|t|)-p\sin(\Omega|t|)\bigr]$, respectively, divided by $\nu$. Their sum is $\frac{e^{-\nu p|t|}}{3\nu p}\bigl[\cos(\Omega|t|)-\frac{p}{2\sqrt{r-p^2}}\sin(\Omega|t|)\bigr]$. Since $\frac{8\gamma T}{9mp}=S(0)$, this result coincides with Eq.~\eqref{eq:Fqt_closed} of the main text. The real-eigenvalue result follows from the same collection of terms or by analytic continuation, $\sqrt{r-p^2}\to i\sqrt{p^2-r}$, $\cos\to\cosh$, $\sin\to i\sinh$, and reads
\begin{equation}
F(\bq,t) = S(0)\; e^{-|t|/\tau(\bq)}\left[ \cosh\!\bigl(\nu\sqrt{p^2-r}\,|t|\bigr) - \frac{p}{2\sqrt{p^2-r}}\,\sinh\!\bigl(\nu\sqrt{p^2-r}\,|t|\bigr) \right].
\label{eq:Fqt_overdamped}
\end{equation}
We have also checked Eqs.~\eqref{eq:Fqt_closed} and \eqref{eq:Fqt_overdamped} numerically against the time correlations of the full linear dynamics at small $q$. Equation~\eqref{eq:Fqt}, with $C^{LL}_\bq(t)$ computed from the full linear dynamics, Eq.~\eqref{eq:timecorr}, has been used in Fig.~\ref{fig:dynamics}(d).

\vskip8pt
\noindent
\textbf{Bragg region --} Away from forward scattering, Eq.~\eqref{eq:FGqt} additionally requires $C^{TT}_\bq(t)$, obtained from $C^{LL}_\bq(t)$ by the replacement $9\to81$ in the $q^4$ prefactor (Sec.~\ref{Timedependentcorrelations}); repeating the same collection of terms gives an expression of the same structural form as Eq.~\eqref{eq:Fqt_closed}, with $S(0)$ replaced by the full angular-dependent static prefactor of Eq.~\eqref{eq:SGq_final} and an amplitude ratio that now differs from $p/(2\sqrt{r-p^2})$ because the $9$ and $81$ prefactors weight the $P_0$ contribution differently between the longitudinal and transverse channels.

\section*{References}
\bibliographystyle{naturemag}
\bibliography{gyroscopicoddsolid}

\end{document}